\documentclass[preprint,aps,amsmath,superscriptaddress,nofootinbib,tightenlines]{revtex4}
\usepackage{graphicx}
\usepackage{bm}
\usepackage{epsfig}

\newif\ifpdf
\ifx\pdfoutput\undefined
\pdffalse 
\else
\pdfoutput=1 
\pdftrue
\fi

\def\xslash#1{{\rlap{$#1$}/}}
\def\Dsl{\hbox{/\kern-.6000em D}} 

\def\dsl{\,\raise.15ex\hbox{/}\mkern-13.5mu D}
\def\bsigma{\mbox{\boldmath $\sigma$}}

\def\psip#1{\psi_{\mathbf{#1}}}
\def\chip#1{\chi_{\mathbf{#1}}}
\def\bsigma{\mbox{\boldmath $\sigma$}}

\def\ltap{\ \raise.3ex\hbox{$<$\kern-.75em\lower1ex\hbox{$\sim$}}\ }
\def\gtap{\ \raise.3ex\hbox{$>$\kern-.75em\lower1ex\hbox{$\sim$}}\ }
\def\OMIT#1{}

\def\lsim{\mathrel{\raise.3ex\hbox{$<$\kern-.75em\lower1ex\hbox{$\sim$}}}}
\def\gsim{\mathrel{\raise.3ex\hbox{$>$\kern-.75em\lower1ex\hbox{$\sim$}}}}

\newcommand{\nn}{\nonumber}

\newcommand{\mNRQCD}{{${\rm mNRQCD}$} }
\newcommand{\mpNRQCD}{mNRQCD--pNRQCD }

\newcommand{\bmk}{\mathbf k}
\newcommand{\bmp}{\mathbf p}
\newcommand{\bmq}{\mathbf q}

\newcommand{\bmA}{\mathbf A}

\newcommand{\bmS}{\mathbf S}

\newcommand{\bmsigma}{\mathbf \bsigma}

\catcode`\@=11
\def\slash{\mathpalette\make@slash}
\def\make@slash#1#2{\setbox\z@\hbox{$#1#2$}%
  \hbox to 0pt{\hss$#1/$\hss\kern-\wd0}\box0}
\catcode`\@=12 

\begin{document}
\ifpdf
\DeclareGraphicsExtensions{.pdf, .jpg}
\else
\DeclareGraphicsExtensions{.eps, .jpg}
\fi


\preprint{ \vbox{ \hbox{MPI-PhT/2002-49} 
\hbox{INT-PUB-02-46}
}}

\title{\phantom{x}\vspace{0.5cm} 
Ultrasoft Renormalization in Non-Relativistic QCD
\vspace{0.5cm} }

\author{Andre H.~Hoang}
\affiliation{Max-Planck-Institut f\"ur Physik
,
F\"ohringer Ring 6, 80805 M\"unchen, Germany
\footnote{Electronic address: ahoang@mppmu.mpg.de}\vspace{0.2cm}}

\author{Iain W. Stewart\vspace{0.4cm}}
\affiliation{Institute for Nuclear Theory,  University of Washington, Seattle, 
	WA 98195 \footnote{Electronic address: iain@phys.washington.edu}
	\vspace{0.5cm}}


\begin{abstract}
\vspace{0.5cm}
\setlength\baselineskip{18pt}

For Non-Relativistic QCD the velocity renormalization group correlates the
renormalization scales for ultrasoft, potential and soft degrees of
freedom. Here we discuss the renormalization of operators by ultrasoft
gluons. We show that renormalization of soft vertices can induce new operators,
and also present a procedure for correctly subtracting divergences in mixed
potential-ultrasoft graphs. Our results affect the running of the
spin-independent potentials in QCD. The change for the NNLL $t\bar t$ cross
section near threshold is very small, being at the $\lesssim 1\%$ level and
essentially independent of the energy. We also discuss implications for
analyzing situations where $mv^2\sim\Lambda_{\rm QCD}$.

\end{abstract}
\maketitle


\newpage

\section{Introduction}

The use of effective field theory techniques in non-relativistic
fermion-antifermion systems has generated an encouraging record of successes.
This framework has applications to describing the dynamics of the bound
quarkonium state as well as to heavy quark production and decay (for reviews see
Ref.~\cite{Reviews}).  The formulation of a non-relativistic field theory for
fermion-antifermion systems in QED and QCD has gone through several stages of
development.  In Refs.~\cite{Caswell,BBL} a method was formulated for separating
the short distance physics at the scale $m$ from the long distance physics at
the non-relativistic momentum and energy scales, $m v$ and $m v^2$.  Subsequent
work served to clarify the power counting in $v$, and the description of the low
energy degrees of freedom. In particular the relevant low energy degrees of
freedom have been classified into potential ($E\sim mv^2$, ${\bf p}\sim mv$),
soft ($E\sim {\bf p}\sim mv$), and ultrasoft ($E\sim {\bf p}\sim mv^2$). 

It was realized in Ref.~\cite{Labelle} that it is necessary to distinguish
ultrasoft and soft contributions, and in Ref.~\cite{LM} that there was a problem
with simultaneously power counting the ultrasoft and potential terms in the
original NRQCD action. In fact, the ultrasoft gluons destroy the power counting
in $v$ unless their Lagrangian is multipole expanded~\cite{GR,Labelle}. This can
be formulated in an elegant way by introducing more than one type of gluon field
in the action~\cite{LS}. Ref.~\cite{PS} made the observation that since
potential gluons do not propagate they can just as well be integrated out of the
theory. In Ref.~\cite{Manohar} it was pointed out that in certain situations the
matching coefficients can be efficiently computed from what is now known as the
hard region of a diagram. This idea was formalized for other momentum regions
and more general situations with the advent of the threshold
expansion~\cite{Beneke}, which also emphasized the importance of soft momenta.
Finally, the relevance of soft gluons for correctly running $\alpha_s$ in the
low energy theory was realized in Ref.~\cite{Gries}.

In Refs.~\cite{PS,PS2,Brambilla} the authors went further to suggest that a
string of effective theories could be defined by exploiting the hierarchy $m\gg
mv\gg mv^2$. We will refer to the theory for scales $mv< \mu < m$ as \mNRQCD,
and for scales $\mu < mv$ as pNRQCD~\cite{PS}.\footnote{Note that we do not use
the term NRQCD to refer to a theory for $\mu>mv$. We reserve NRQCD as a generic
label for the effective theory(s) which describe $c\bar c$, $b\bar b$, and
$t\bar t$ bound state effects.}  A correct separation of the $mv$ and $mv^2$
scales is important since we would like to be confident about what effects can
be reliably computed perturbatively, for instance in the case
$mv^2\sim\Lambda_{\rm QCD}\ll mv$.  The \mpNRQCD setup appears to rely on being
able to treat the energy and momentum scales of the non-relativistic quarks and
their corresponding cutoff scales as independent. For static heavy quarks this
is the case because the dynamics does not correlate the quark
separation $r\sim 1/(mv)$ with energy fluctuations $E\ll mv$.

In Ref.~\cite{LMR} it was pointed out that for dynamic heavy quarks the
non-relativistic dispersion relation $E=p^2/(2m)$ couples the energy and
momentum scales. The authors therefore proposed matching directly at the scale
$\mu=m$ onto a potential-like theory with both soft and ultrasoft degrees of
freedom (referred to as vNRQCD). For scales $\mu< m$ the correlation of energy
and momenta is accounted for since the ultrasoft and soft scales, $\mu_U$ and
$\mu_S$, are related, $\mu_U=\mu_S^2/m\equiv m\nu^2$. The running in the
dimensionless velocity parameter from $\nu=1$ to $\nu\simeq v_0$ of order the
velocity of the two particle state, sums logs of both the momenta and the energy
at the same time, and is referred to as the ``velocity Renormalization Group''
(vRGE)~\cite{LMR,amis}. Within dimensional regularization the factors of
$\mu_U^\epsilon$ or $\mu_S^\epsilon$ multiplying operators in the renormalized
effective Lagrangian are uniquely determined from the $v$ power counting in $d$
dimensions~\cite{amis2}. This point is also discussed in
Sect.~\ref{sect_formalism}.

With the vRGE, the running of potentials and currents was worked out in
Refs.~\cite{LMR,amis,amis3,hms1} and applied to $t\bar t$ production near
threshold in Refs.~\cite{hmst,hmst1}. The running of the static potential due to
ultrasoft effects in the \mpNRQCD formalism was computed in
Ref.~\cite{PSstat}. In Ref.~\cite{amis4} it was shown that the vRGE could be
used to predict $\ln\alpha$ contributions in QED bound states, and for
positronium the $\alpha^7\ln^2\alpha$ hyperfine splitting corrections and
$\alpha^3\ln^2\alpha$ corrections to decay rates were reproduced, and the
$\alpha^8\ln^3\alpha$ Lamb shift was predicted. In Ref.~\cite{mss1} it was shown
that the correlation of energy and momentum scales is necessary to compute QED
corrections involving $\ln^k\alpha$ with $k\ge 2$.  More recently, the running
of operators in the \mpNRQCD framework were computed~\cite{Pineda1}, and the
original formalism was modified in Ref.~\cite{Pineda2} to include the
correlation of potential and ultrasoft cutoffs.

The main purpose of this paper is to discuss two aspects of the ultrasoft
renormalization of operators for $mv^2\gg\Lambda_{\rm QCD}$. We first point out
a new set of soft operators which has zero tree-level matching, but is induced
by mixing from ultrasoft renormalization of soft time-ordered products. These
operators vanish for QED bound states such as Hydrogen or positronium. However
they affect the leading logarithmic running of the two spin-independent $1/m^2$
QCD potentials as discussed in Sec.\,~\ref{sectionQCDm2}, and were not included
in Ref.~\cite{amis}. We also compare our results to those in Ref.~\cite{Pineda1}
where analogous operators were included. In Ref.~\cite{Pineda1} two results were
reported, one for scales $mv< \mu < m$ (\mNRQCD) and one for $\mu< mv$
(pNRQCD). For the $1/m^2$ potentials we disagree with the \mNRQCD results
because we find that for dynamic quarks there is no corresponding momentum
region we can identify since the renormalization from ultrasoft gluons is always
present. For the pNRQCD results we find agreement, however only if we force the
\mpNRQCD matching scale and the pNRQCD energy cutoffs to always be correlated.
We also show that our approach reproduces the $\alpha^6\ln^2\alpha$ energy for
muonic-Hydrogen with additional massless fermions~\cite{Pineda3}.

Second, we formulate a procedure for correctly subtracting ultrasoft divergences
in diagrams containing both ultrasoft and potential loops.  Our results for the
running of the $1/(m|\bmk|)$ and $1/\bmk^2$ QCD potentials differ from
Refs.~\cite{amis3,PSstat,hms1,Pineda1}, essentially because we find that there
is an additional set of operators ${\cal O}_{ki}$, ${\cal O}_{ci}$ which must be
renormalized. For matrix elements which do not involve additional divergences,
the sum of these operators reduce to effective $1/(m|\bmk|)$ and $1/\bmk^2$
potentials. For these effective potentials we agree with the $\mu<mv$ pNRQCD
results of Ref.~\cite{Pineda1} (again only if we demand that there is always a
correlation between the \mpNRQCD matching scale and energy cutoffs). However,
our results also apply to matrix elements with additional divergences such as
corrections to the $Q\bar Q$ current correlators $G(0,0)$, where effective
$1/(m|\bmk|)$ and $1/\bmk^2$ potentials do not suffice.  Numerically the change
from the earlier QCD results is quite small, being $\lesssim 1\%$ for the
normalization of the $t\bar t$ cross section. Our results have implications for
the less perturbative situation $mv^2\sim\Lambda_{\rm QCD}$ for dynamic
quarks. They imply that NRQCD in this situation may be more non-perturbative
than sometimes assumed.

In Sec.~\ref{sect_formalism} we review some of the formalism that we will need.
We outline the issues associated with the ultrasoft renormalization of operators
in Sect.~\ref{sect_usoft} as well as our solutions. This is applied to muonic
Hydrogen in Sec.\,\ref{sectionmH}, the $1/m^2$ QCD potentials in
Sec.\,\ref{sectionQCDm2}, and the effect on the $1/(m|\bmk|)$ and $1/\bmk^2$ QCD
potentials is taken up in Sec.\,\ref{sectionQCDm0m1}. We discuss the running of
the production current in Sec.~\ref{sectionc1}, as well as implications for the
scenario where $mv^2\sim\Lambda_{\rm QCD}$. In Sec.\,~\ref{sectiontop} we
discuss results for the production current correlators and give numerical
results for the $e^+e^-\to t\bar t$ cross section near threshold. Conclusions
are given in Sec.\,\ref{section_conclusion}.

\section{Formalism} \label{sect_formalism}

We begin be reviewing some aspects of the NRQCD formalism used here. As a way of
motivation it is useful to recall that a properly constructed effective theory
should satisfy the following requirements:
\begin{enumerate}
 \item reproduce the IR divergences of the full theory in its entire region 
   of validity, 
 \item have a well defined power counting (in $v$ for our case), 
 \item have no large logs in matching calculations,
 \item start with a regulator independent Lagrangian.
\end{enumerate}
The first property ensures that we have included the correct degrees of freedom.
The need for the second property is obvious. The third property follows from the
first, and the matching will be independent of the IR regulator as long as the
same choice is made in the full and effective theories. The fourth property is
necessary so that physical predictions are independent of the method used to
regulate ultraviolet divergences.  In practice we choose dimensional
regularization which makes it easy to preserve the symmetries and power
counting. For this case the fourth constraint dictates that the action is
independent of $d$ when expressed in terms of bare quantities.

The physical system we wish to describe is that of a heavy fermion and
antifermion with mass $m$, and energies $E\sim mv^2$, and momenta ${\bf p}\sim
mv$ in the c.m.\,system where $v\ll 1$. The possible degrees of freedom
include~\cite{Caswell,BBL,Labelle,LM,GR,LS,PS,Manohar,Beneke,Gries} heavy
potential quarks and antiquarks ($\psi_\bmp$, $\chi_\bmp$), ultrasoft gluons,
ghosts, and massless quarks ($A^\mu$, $c$, $\varphi_{us}$), soft gluons, ghosts,
and massless quarks ($A_q^\mu$, $c_q$, $\varphi_q$), potential gluons ($A_{\rm
p}^\mu$) and soft heavy quarks and antiquarks ($\psi^s_q$, $\chi^s_q$). Here,
ultrasoft gluons are the gauge partners of momenta $\sim mv^2$, while soft
gluons are the partners of momenta $\sim mv$.  It is essential that we include
both soft and ultrasoft gluons for all scales less than $m$.  Evaluating a
generic QCD scattering amplitude in the region of validity of the effective
theory gives logarithms
\begin{eqnarray}
  \ln(E^2)\,,\qquad \ln({\bf p}^2)\,,\qquad \ln({\bf k}^2) \,,
\end{eqnarray}
where $E$ denotes the c.m.\,energy, $\bmp$ a quark momentum, and ${\bf k}$ the
momentum transfer. Both ultrasoft and soft gluons are needed to reproduce the
$\ln(E^2)$ and $\ln({\bf k}^2)$ terms.  Furthermore, both types of logarithm are
needed for all scales below $m$ since both ultrasoft and soft running feed into
anomalous dimensions induced by potential loops such as the production
current~\cite{LMR} and two-loop renormalization of $1/m^2$ operators in
QED~\cite{amis4}. Simultaneously including soft and ultrasoft gluons is also in
agreement with the threshold expansion~\cite{Beneke} where both soft and
ultrasoft regions of energy and momenta are included in calculations in a
separated form at scales $\mu\simeq m$.  When these degrees of freedom are not
treated separately, such as in the mNRQCD-pNRQCD approach for $mv<\mu<m$, the $v$
power counting breaks down, and a $1/m$ expansion with static quarks is
used~\cite{Pineda1}. However, it is argued that a power counting for dynamic
quarks exists in pNRQCD for scales $\mu <mv$~\cite{Pineda1}.

To distinguish the $mv$ and $mv^2$ scales we made use of a phase redefinition
for the potential and soft fields~\cite{LMR}
\begin{eqnarray}
  \phi(x) \to \sum_k e^{-ik\cdot x} \phi_k(x) \,,
\end{eqnarray}
where $k$ denotes momenta $\sim mv$ and $\partial^\mu \phi_k(x)\sim mv^2
\phi_k(x)$.  Since here we will only be interested in cases with perturbative
potentials (such as $\Lambda_{\rm QCD}\ll mv^2$) we simplify the list of degrees
of freedom by integrating out potential gluon exchange and soft heavy fermions
at the scale $m$ following Ref.~\cite{LMR}. This choice may not be unique, but
does allow us to satisfy our effective theory criteria. For instance, in
Ref.~\cite{amis2} it was shown that vNRQCD correctly reproduces
all the infrared logs in QCD for four-quark Greens functions at one
loop in its entire region of validity.

The effective vNRQCD Lagrangian can be separated into ultrasoft, soft, and
potential components, ${\cal L}={\cal L}_u + {\cal L}_s + {\cal L}_p$.  The
presence of two types of gluons immediately brings up the issue of double
counting.  To avoid double counting the effective theory is constructed such
that the ultrasoft gluons reproduce only the physical gluon poles where $k^0\sim
{\bmk}\sim mv^2$, while soft gluons give only those with $k^0\sim {\bmk}\sim
mv$. The scales for the gluon momenta are influenced by the quark propagators,
so the quark-gluon interactions must be constructed in such a way that we will
not upset this scaling.  In ${\cal L}_u$ this is achieved by the multipole
expansion of interactions~\cite{Labelle,GR}, which ensures that ultrasoft gluon
momenta are always much smaller than the quark three-momenta.  The ultrasoft
Lagrangian is~\cite{GR,LMR}
\begin{eqnarray} \label{Lus}
{\cal L}_u  &=& 
\sum_{\mathbf p}\,\bigg\{
   \psi_{\bmp}^\dagger   \bigg[ i D^0 - \frac {\left({\bf p}-i{\bf D}\right)^2}
   {2 m} +\frac{{\mathbf p}^4}{8m^3} + \ldots \bigg] \psi_{\bmp}
 + (\psi \to \chi)\,\bigg\}
 -\frac{1}{4}G_u^{\mu\nu}G^u_{\mu \nu}\nn\\
 && +\ldots \,,
\end{eqnarray}
where $G_u^{\mu\nu}$ is the ultrasoft field strength. In dimensional
regularization the covariant derivative has the form $D^\mu=\partial^\mu+i
\mu_U^\epsilon\, g_u A^\mu$, where $\mu_U=m\nu^2$ and $g_u=g_u(\mu_U)$ is the
renormalized ultrasoft QCD coupling.  Note that the covariant derivative only
contains the ultrasoft gluon field and that the ultrasoft Lagrangian has the
form of the multipole-expanded HQET Lagrangian. For convenience we suppress
(throughout this paper) the renormalization $Z$ factors that relate bare and
renormalized quantities. The soft Lagrangian has terms~\cite{LMR,amis,amis2}
\begin{eqnarray} \label{Ls}
{\cal L}_s &=& 
  \sum_q \Big\{ \bar\varphi_q\: \xslash{q}\: \varphi_q 
  -\frac{1}{4}G_s^{\mu\nu} G^s_{\mu \nu}
  + \bar c_q q^2 c_q\Big\}
  \\
  &&
  - g_s^2 \mu_S^{2\epsilon}\! 
  \sum_{{\bmp},{\bmp^\prime},q,q^\prime,\sigma} \bigg\{ 
  \frac{1}{2}\, \psi_{\bmp^\prime}^\dagger
  [A^\mu_{q^\prime},A^\nu_{q}] U_{\mu\nu}^{(\sigma)} \psi_{\bmp}
 + \frac{1}{2}\,
 \psi_{\bmp^\prime} \{A^\mu_{q^\prime},A^\nu_{q}\} W_{\mu\nu}^{(\sigma)}
 \psi_{\bmp}
 \nn \\[2mm]\
 &&+ {\psip {p^\prime}}^\dagger\: [\bar c_{q'}, c_q] Y^{(\sigma)}\:
 {\psip p}
 + ({\psip {p^\prime}}^\dagger\: T^B Z_\mu^{(\sigma)}\:
 {\psip p} ) \:(\bar \varphi_{q'} \gamma^\mu T^B \varphi_q)  \bigg\}
 + (\psi \to \chi,\: T\to \bar T) \,,\nn 
\end{eqnarray}
where $G_s^{\mu\nu}$ is the soft gluon field strength and $g_s=g_s(\mu_S)$
($\mu_S=m\nu$) is the soft QCD coupling.  The tensors $U_{\mu\nu}^{(\sigma)}$,
$W_{\mu\nu}^{(\sigma)}$, $Z_\mu^{(\sigma)}$, and $Y^{(\sigma)}$ are functions of
${\bf p'}, {\bf p}, q, q'$ that are generated by integrating out soft heavy
quarks and potential gluons. Their explicit form can be found in
Ref.~\cite{amis3}. Finally the potential Lagrangian has terms~\cite{LMR,amis3}
\begin{eqnarray} \label{Lp}
{\cal L}_p &=& -\mu_S^{2\epsilon}\, V({\bmp,\bmp^\prime})\,
   \psi_{\bmp^\prime}^\dagger \psi_{\bmp}
   \chi_{-\bmp^\prime}^\dagger \chi_{-\bmp}
   + \mu_S^{2\epsilon} F_j^{ABC}({\bf p},{\bf p'})(g_u\mu_U^\epsilon {\bmA}^C_j) 
   \Big[\psi_{\bmp^\prime}^\dagger T^A \psi_{\bmp} 
   \chi_{-\bmp^\prime}^\dagger \bar T^B \chi_{-\bmp} \Big] \nn \\
 && + \ldots \,,
\label{vNRQCDLagrangian}
\end{eqnarray}  
where $\mu_S^{2\epsilon}V$ and $\mu_S^{2\epsilon}F_j^{ABC}$ are functions
involving bare Wilson coefficients. In the first term spin and color indices on
$V$ and the fermion fields have been suppressed. Matching perturbatively at $m$
and integrating out the potential gluons generates the terms
\begin{eqnarray}
 && V({\bmp},{\bmp^\prime}) =   (T^A \otimes \bar T^A) \bigg[
 \frac{{\cal V}_c^{(T)}}{\bmk^2}
 + \frac{{\cal V}_k^{(T)}\pi^2}{m|{\bmk}|}
 + \frac{{\cal V}_r^{(T)}({\bmp^2 + \bmp^{\prime 2}})}{2 m^2 \bmk^2}
 + \frac{{\cal V}_2^{(T)}}{m^2}
 + \frac{{\cal V}_s^{(T)}}{m^2}{\bmS^2} \nn \\[2mm] 
&&\quad + \frac{{\cal V}_\Lambda^{(T)}}{m^2}\Lambda({\bmp^\prime ,\bmp}) 
 + \frac{{\cal V}_t^{(T)}}{m^2}\,
 T({\bmk}) + \ldots \bigg] 
 + (1\otimes 1)\bigg[
 \frac{{\cal V}_c^{(1)}}{\bmk^2}
 + \frac{{\cal V}_k^{(1)}\pi^2}{m|{\bmk}|}
 + \frac{{\cal V}_2^{(1)}}{m^2}
 + \frac{{\cal V}_s^{(1)}}{m^2}{\bmS^2} +\ldots \bigg]
\,, \nonumber \\[2mm]
 &&\quad {\bmS} \, = \, \frac{ {{\bmsigma}_1 + {\bmsigma}_2} }{2}\,,
 \quad 
 \Lambda({\bmp^\prime},{\bmp}) \, = \, -i \frac{{\bmS} \cdot ( {\bmp^\prime}
 \times {\bmp}) }{  {\bmk}^2 }\,,\quad
 T({\bmk}) \, = \, {{\bmsigma}_1 \cdot {\bmsigma}_2} - \frac{3\, {{\bmk}
 \cdot {\bmsigma}_1}\,  {{\bmk} \cdot {\bmsigma}_2} }{ {\bmk}^2} \,,
 \nn\\[2mm]
 && F^{ABC}_j({\bmp},{\bmp^\prime}) = \frac{2i{\cal V}_c^{(T)}\bmk_j }{\bmk^4} 
 f^{ABC} \,,
\label{vNRQCDpotential}
\end{eqnarray}
where $\bmk=\bmp'-\bmp$. The factors ${\cal V}_j(\nu)$ are renormalized Wilson
coefficients, and the coefficient ${\cal V}_c^{(T)}$ in $F_j^{ABC}$ is fixed by
reparameterization invariance~\cite{LM} as shown in Ref.~\cite{amis3}.  Some
further operators are required to renormalize ${\cal L}_{s,p}$ and will be
discussed later on.

It is worth noting that the factors of $\mu_U^\epsilon$ and $\mu_S^\epsilon$ in
${\cal L}_{u,s,p}$ are uniquely determined by mass dimension and $v$ power
counting in $d=4-2\epsilon$ dimensions~\cite{amis2}. A scaling with $v$ is
assigned to the effective theory fields so that in the action their kinetic
terms are order $v^0$. This gives $\psi_{\bf p}\sim (mv)^{3/2-\epsilon}$,
$A^\mu\sim (mv^2)^{1-\epsilon}$, and $A_q^\mu\sim (mv)^{1-\epsilon}$. Since
$D^\mu\sim mv^2$, the renormalized combination $g_u A^\mu$ must be multiplied by
$\mu_U^\epsilon \sim (mv)^{\epsilon}$ so that this gluon term also scales as
$mv^2$. For the potential fermion terms in ${\cal L}_s$ displayed in
Eq.~(\ref{Ls}) the soft and potential fields give a $(mv)^{5-4\epsilon}$ and the
complete measure gives $(mv)^{-4+2\epsilon}$. The factor of
$\mu_S^{2\epsilon}\sim (mv)^{2\epsilon}$ is therefore required to cancel the
$v^{-2\epsilon}$ from the measure and fields.  For the four-quark operator in
${\cal L}_p$ the quark fields give $(mv)^{6-4\epsilon}$ and the complete measure
gives $(mv^2)^{-1} (mv)^{-3+2\epsilon}$, so a factor of $\mu_S^{2\epsilon}\sim
(mv)^{2\epsilon}$ multiplying $V({\bf p,p'})$ is required. The factors of
$\mu_U^\epsilon$ and $\mu_S^\epsilon$ for any other operator can also be
determined in this way.

The power counting of an arbitrary diagram is determined entirely by the powers
of $v$ assigned to its operators. Since the factors of $\mu_{U}^\epsilon$ and
$\mu_S^\epsilon$ are already determined, we can work in $d=4$ dimensions.  In
general a graph is order $\sim v^\delta$ with~\cite{LMR}
\begin{eqnarray} \label{pc}
 \delta = 5+  \sum_k \Big[\, 
 (k-8) V_k^{(u)} + (k-5) V_k^{(p)} + (k-4) V_k^{(s)}
\,\Big] -N_s\,.
  \\[-5pt]\nn
\end{eqnarray} 
Here $N_s$ is the number of disconnected soft subgraphs if potential lines are
erased, the vertex factors $V_k^{(i)}$ count the number of insertions of an
operator scaling as $v^k$. The factors $V_k^{(u)}$ are for
purely ultrasoft operators, $V_k^{(p)}$ are for operators which contain $\psip p$
or $\chip p$ but no soft fields, and $V_k^{(s)}$ are for operators
with at least one soft field. In this paper we will refer to the order in $v$ of
diagrams and operators as their value of
\begin{eqnarray}
 \delta' = \delta-5 \,,
\end{eqnarray}
since this quantity is essentially additive with multiple insertions of
operators~\cite{amis4}. For instance, using the value of $\delta'$ for the power
counting a T-product of the potentials $V_1\sim v^a$ and $V_2\sim v^b$ scales as
$T[V_1 V_2]\sim v^{a+b}$.  With this counting the Coulomb potential scales as
$v^{-1}$, the $1/|{\bf k}|$ potential is order $v^0$ and the $1/m^2$ potentials
are order $v^1$. For later use we also define
\begin{equation}
\label{sigmadef}
\sigma=k-4
\end{equation}
as the order in $v$ for which a soft operator contributes to the power
counting formula in Eq.\,(\ref{pc}).

\section{Ultrasoft Renormalization of Operators} \label{sect_usoft}

Since the effective theory simultaneously involves soft and ultrasoft gluons the
question naturally arises in what manner the two types of gluons renormalize
operators.  As far as renormalization of ${\cal L}_u$ is concerned the ultrasoft
renormalization is equivalent to that in HQET. This is because after the
equation of motion $E={\bf p^2}/(2m)$ is applied, the quark propagators in loop
graphs in the single fermion sector become static~\cite{LMR},
\begin{eqnarray}
  \frac{i}{E+k^0-{\bf p^2}/{(2m)}+i\epsilon} = \frac{i}{k^0+i\epsilon} \,.
\end{eqnarray}
Here ($E$, ${\bf p}$) are the energy and momentum of the external quark line.
Thus, the renormalized ${\cal L}_u$ is exactly of the form of the
multipole-expanded renormalized HQET Lagrangian.

The ultrasoft renormalization of interactions generated by ${\cal L}_s$ and
${\cal L}_p$ is more subtle. The basic complication is that one must account
for the fact that ultrasoft gluons renormalize operators which are
non-local with respect to the $mv$ scale (but are local relative to $mv^2$).

\begin{figure}[t!]
\centerline{ 
\includegraphics[width=1.5in]{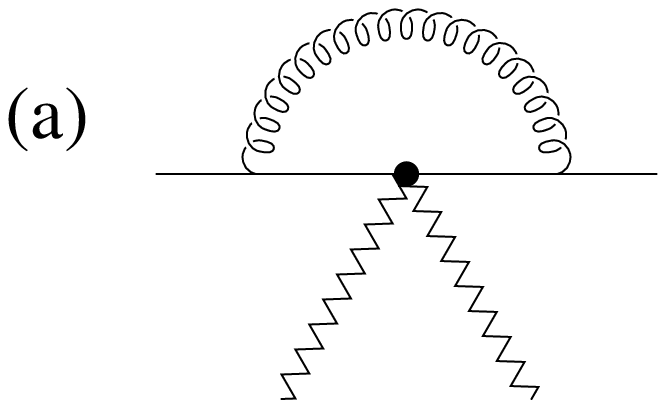}\hspace{1.7cm} 
\includegraphics[width=1.7in]{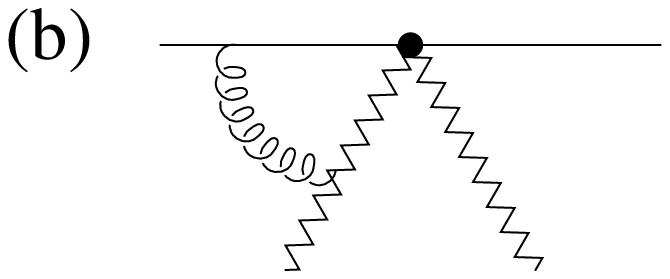} 
 }
\caption{Examples of mixed soft-ultrasoft graphs with $A^0$ ultrasoft gluons
which could renormalize ${\cal L}_s$ in Feynman gauge. The zigzag lines denote
soft gluons, quarks, or ghosts.\label{fig_gdep}}
\end{figure}

First consider the renormalization of interactions generated by ${\cal L}_s$.
Directly renormalizing ${\cal L}_s$ by ultrasoft gluons leads to gauge dependent
Wilson coefficients as observed in Ref.~\cite{amis3}. To see this consider the
graphs in Fig.~\ref{fig_gdep} which involve ultrasoft $A^0$ gluons.
In Coulomb gauge these graphs are identically zero. In Feynman gauge the sum of
Fig.~\ref{fig_gdep}a and the quark wavefunction renormalization give a
$C_A/\epsilon$ term which could contribute to the renormalization of
$U^{(0)}_{\mu\nu}$ in ${\cal L}_s$.  Non-Abelian graphs such as the one in
Fig.~\ref{fig_gdep}b do not cancel this contribution. For Fig.~\ref{fig_gdep}b,
the intermediate soft gluon propagator has momentum $(q+k)^2 = q^2 + 2k\cdot
q+k^2$, but we cannot set the offshellness $q^2\sim (mv)^2$ of the soft gluon
to zero without violating the power counting and risking a double
counting. Keeping $q^2\ne 0$, Fig.~\ref{fig_gdep}b evaluates to zero and it is
not surprising that the total result is gauge dependent since part of the
calculation was done offshell. This issue can be avoided by only considering the
ultrasoft renormalization of operators which can contribute as soft color
singlets~\cite{amis3}, such as time-ordered products involving two or more soft
vertices and quarks and antiquarks, $T({\cal L}_s^\psi\, {\cal L}_s^\chi \,
\ldots)$. These products appear local as far as the ultrasoft gluon is
concerned, and it is only these products which affect observables. Having
ultrasoft renormalization only for these operators also avoids the predicament
of having both ultrasoft and soft gluons in the single heavy quark sector. Thus,
we do not consider ultrasoft renormalization of a single ${\cal L}_s$ term.

\begin{figure}[t!]
\centerline{ 
\includegraphics[width=3in]{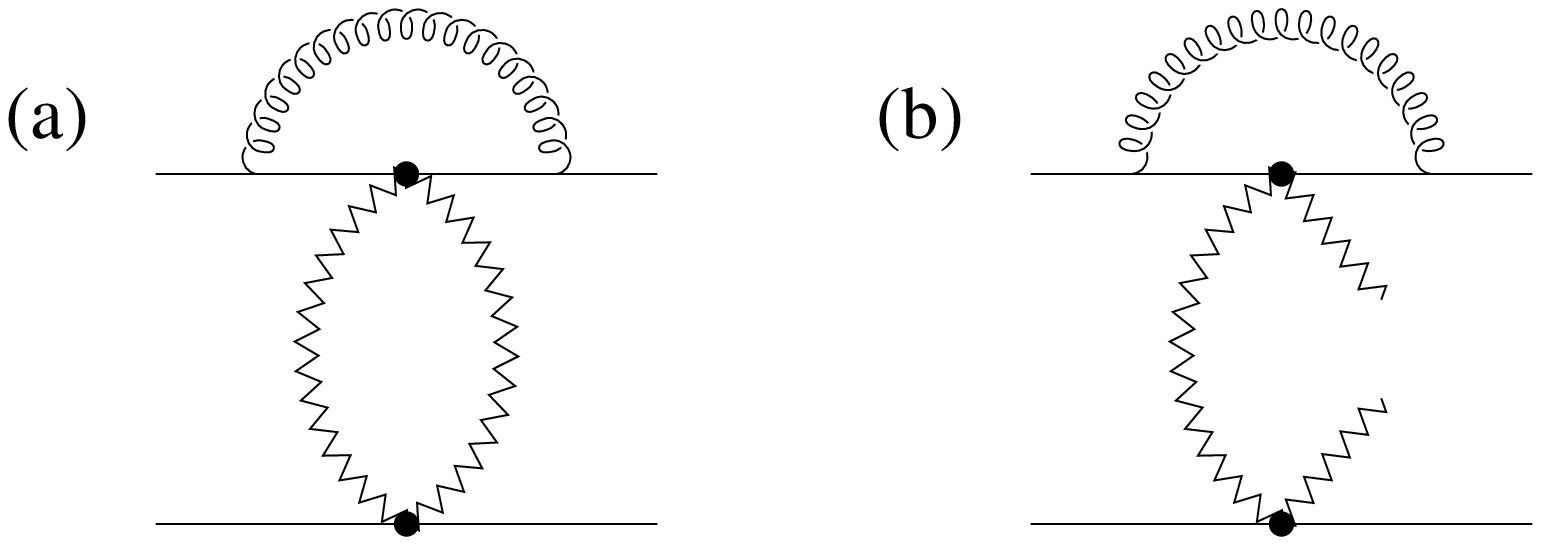}\hspace{0.7cm} 
\includegraphics[width=1.7in]{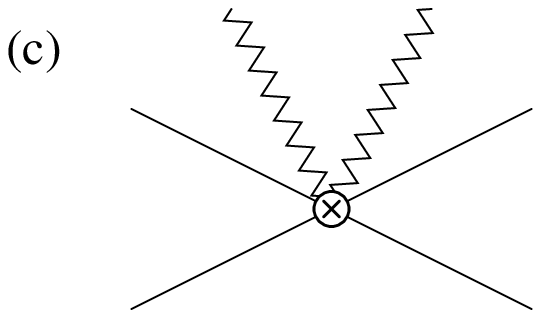} 
 }
\caption{The zigzag lines denote soft gluons, quarks, or ghosts. Graphs (a) and
(b) are examples of mixed soft-ultrasoft graphs, while (c) denotes an operator
for soft Compton scattering off a potential.
\label{fig_sus}}
\end{figure}

\begin{figure}[t!]
\centerline{ 
\includegraphics[width=1.38in]{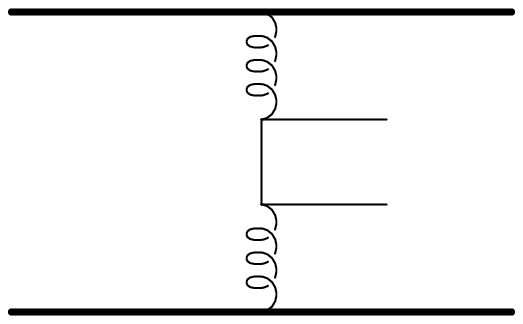}\hspace{0.5cm} 
\raisebox{1cm}{---} \hspace{0.5cm} 
\includegraphics[width=1.2in]{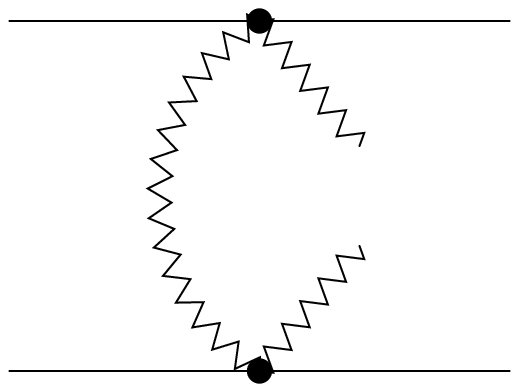}\hspace{0.5cm} 
\raisebox{1cm}{=} \hspace{0.7cm}  
\includegraphics[width=1.2in]{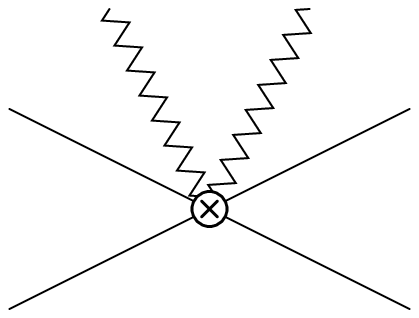} 
 }
\caption{Example of the matching calculation for ${\cal
O}_{2\varphi}^{(\sigma)}$. Here the zig-zag lines denote soft massless
fermions. At the high scale ($\nu=1$) the graphs on the left exactly cancel so
the coefficient of the operator on the right is zero.
\label{fig_sm}}
\end{figure}

However this does {\em not} imply that the first ultrasoft renormalization of a
graph involving soft vertices are for two loops graphs like the one in
Fig.~\ref{fig_sus}a.
The reason is that we must first consider the case where only one pair of soft
gluon fields in the time-ordered product are contracted. So at one-loop we must
consider the renormalization of graphs such as the one shown in
Fig.~\ref{fig_sus}b.  The ultraviolet divergences in these graphs need to be
canceled by counterterms for the 6-field operators, ${\cal O}_{2i}^{(\sigma)}$
shown in Fig.~\ref{fig_sus}c. For quarks, gluons, and ghosts these operators
have the structure
\begin{eqnarray} \label{6q}
  {\cal O}_{2\varphi}^{(\sigma)} &=& { g_s^4\,\mu_S^{4\epsilon}}\:  \:
   (\psi_{\bmp^\prime}^\dagger\: \Gamma^{(\sigma)}_{\varphi,\psi}\: 
   \psi_{\bmp}) \:
   (\chi_{-\bmp^\prime}^\dagger\: \Gamma^{(\sigma)}_{\varphi,\chi}\: 
   \chi_{-\bmp})
   \ (\overline\varphi_{-q}\: \Gamma_\varphi^{(\sigma)}\: \varphi_{q}) \,, \nn\\
 {\cal O}_{2A}^{(\sigma)}  &=& { g_s^4\,\mu_S^{4\epsilon}}\:  \:
   (\psi_{\bmp^\prime}^\dagger\: \Gamma^{(\sigma)}_{A,\psi} \: \psi_{\bmp})  \:
   (\chi_{-\bmp^\prime}^\dagger\: \Gamma^{(\sigma)}_{A,\chi} \: \chi_{-\bmp})
   \ (A^\mu_{-q}\: \Gamma_{A,\mu\nu}^{(\sigma)}\: A^\nu_{q}) \,,\\\
 {\cal O}_{2c}^{(\sigma)}  &=& { g_s^4\,\mu_S^{4\epsilon}}\:  \:
   (\psi_{\bmp^\prime}^\dagger\: \Gamma^{(\sigma)}_{c,\psi} \: \psi_{\bmp})  \:
   (\chi_{-\bmp^\prime}^\dagger\: \Gamma^{(\sigma)}_{c,\chi} \: \chi_{-\bmp})
   \ (\bar c_{-q}\: \Gamma_c^{(\sigma)}\: c_{q}) \,,\nn
\end{eqnarray}
where color indices are suppressed and the factor of $\mu_S^{4\epsilon}$ is
determined by the procedure in Sect.~\ref{sect_formalism}.  Here the $\Gamma_i$
are matrices in spin and/or color space and can be functions of ${\bf p}', {\bf
p}$, and $q^\mu$. This soft momentum dependence is identical to that in the
graph on the LHS of Fig.~\ref{fig_sm} once the equations of motion are applied
(e.g. $q^2=0$, $\xslash{q} \varphi_{q}=0$).  The superscript $\sigma\ge 0$ has
been used in Eq.\,(\ref{sigmadef}) and denotes the order in $v$ for which these
soft operators contribute to the power counting in Eq.\,(\ref{pc}).  For our
purposes the $\sigma=0,2$ operators will be sufficient. The operators in
Eq.~(\ref{6q}) also have Wilson coefficients $C_{2i}^{(\sigma)}(\nu)$. The
tricky thing about the operators ${\cal O}_{2i}^{(\sigma)}$ is that the tree
level matching onto their Wilson coefficients is zero,
$C_{2i}^{(\sigma)}(\nu=1)=0$. An example of this matching calculation is shown
in Fig.~\ref{fig_sm} for ${\cal O}_{2\varphi}^{(\sigma)}$.  At $\nu=1$ the full
theory graph on the left is exactly canceled by the time ordered product of soft
vertices. Therefore, we get zero for the Wilson coefficient of the operator on
the right. However, the ultrasoft loop graph in Fig.~\ref{fig_sus}b gives ${\cal
O}_{2i}^{(2)}$ a non-zero anomalous dimension so that for $\nu<1$ the
coefficient evolves and $C_{2i}^{(2)}(\nu<1)\ne 0$. The same counterterm is
necessary to renormalize the divergences in Fig~\ref{fig_sus}a. Furthermore, the
operators ${\cal O}_{2i}^{(2)}$ affect the anomalous dimensions for the $1/m^2$
coefficients ${\cal V}_2$ and ${\cal V}_r$ in Eq.~(\ref{vNRQCDpotential}), and
were not included in Ref.~\cite{amis}. This occurs through an ultraviolet
divergence in the graph in Fig.~\ref{fig_s3}c which needs to be canceled by
${\cal V}_{2,r}$ counterterms.  In Secs.~\ref{sectionmH} and \ref{sectionQCDm2}
we will give two examples of the implications of the operators in
Eq.~(\ref{6q}).

\begin{figure}[t!]
\centerline{ 
\includegraphics[width=1.5in]{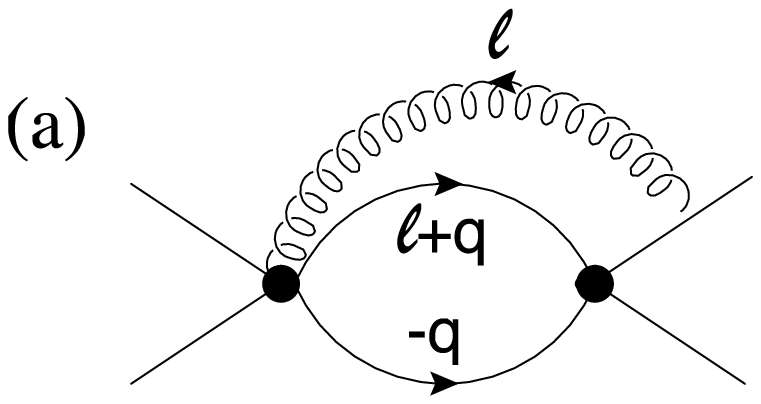} \hspace{1.5cm}
\includegraphics[width=1.5in]{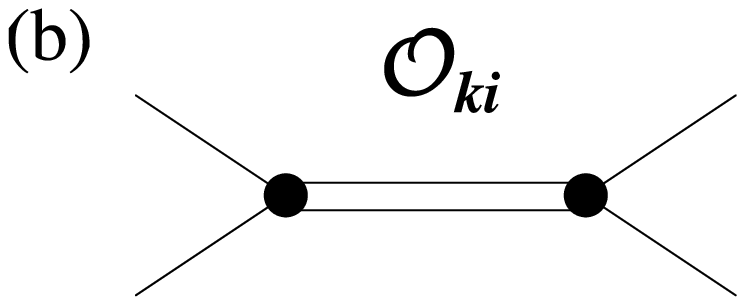}   }
\vspace{0.3cm}
\centerline{ 
\includegraphics[width=1.5in]{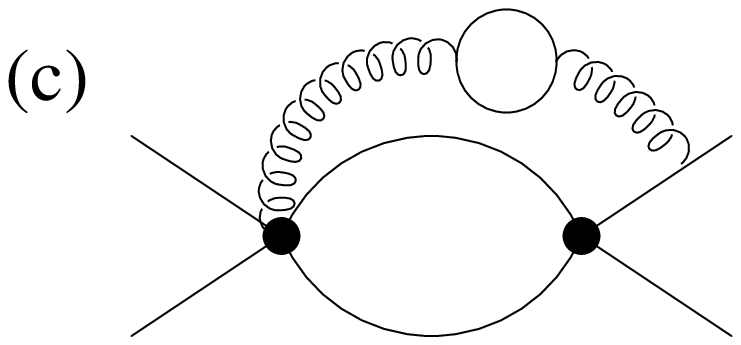} \hspace{1.5cm}
\includegraphics[width=1.5in]{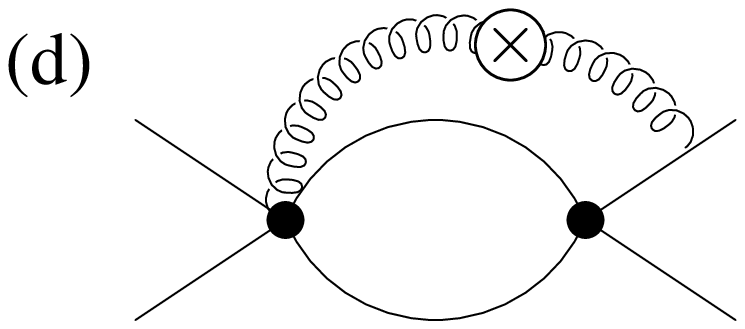} \hspace{1.5cm}
\includegraphics[width=1.5in]{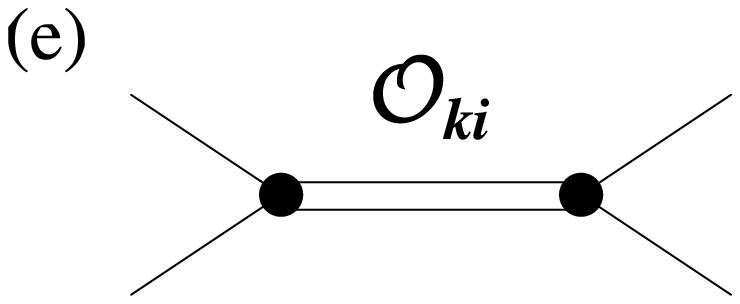}   }
\vspace{-0cm} \caption{Example of renormalization of two-body potential 
graphs by an ultrasoft gluon. The ultrasoft couplings are the order $v^0$
term from Eq.~(\ref{Lp}) and the order $v^1$ term ${\bf p}\cdot {\bf A}$ in
Eq.~(\ref{Lus}).
\label{fig_us1}}
\end{figure}

Let us now discuss the ultrasoft renormalization for two-body interactions
generated by ${\cal L}_p$. In general in graphs with two heavy quarks the
equations of motion do not make the propagators static. For example,
in the order $v^0$ two-loop graph in Fig.~\ref{fig_us1}a the ultrasoft
loop momentum $\ell^\mu$ is routed through an internal fermion line
giving a propagator 
\begin{eqnarray}
  \frac{i}{E+\ell^0 + q^0 - {\bf q}^2/(2m)+i\epsilon} 
  =\frac{i}{\ell^0 + q^0 - ({\bf q}^2\!-\!{\bf p}^2)/(2m) + i\epsilon}\,,
\end{eqnarray}
where $q=(q^0,{\bf q})$ is the potential loop momentum.  The multipole expansion
has resulted in factors of the three-momentum {\boldmath $\ell$} being dropped,
however the remaining propagator is not static in general since ${\bmq}^2\ne
{\bf p}^2$.  In fact, the ultrasoft loop induces an UV divergence after the
$dq^0$ and $d^d\ell$ integrals have been performed, but before the sum over
indices ${\bf q}$ is carried out.
In Refs.~\cite{amis3,hms1} the subtraction of the ultrasoft UV divergences was
made after carrying out the sum over ${\bf q}$.  Instead, the corresponding
operators used to subtract the divergences should involve a sum over $\bmq$. For
example, for the sum of all scattering diagrams at order $\alpha_s^3 v^0$ with
one ultrasoft gluon the operators needed to subtract the ultrasoft UV
divergences have the form
\begin{eqnarray} \label{Ok1a}
{\cal O}_{k1}^{(1)} &=& - \frac{[\mu_S^{2\epsilon}\,{\cal V}_c^{(T)}]^2}{m}\: 
 \sum_{{\bf p,p',q}} ( f_{0} + f_{1} + 2 f_{2} )\
 \big[ \psi_{\bmp^\prime}^\dagger \psi_{\bmp}
 \chi_{-\bmp^\prime}^\dagger \chi_{-\bmp} \big] \,, \nn\\
{\cal O}_{k2}^{(T)} &=& - \frac{[\mu_S^{2\epsilon}\,{\cal V}_c^{(T)}]^2}{m}\: 
 \sum_{{\bf p,p',q}} ( f_{1} + f_{2} ) \ 
 \big[ \psi_{\bmp^\prime}^\dagger T^A \psi_{\bmp}
 \chi_{-\bmp^\prime}^\dagger \bar T^A \chi_{-\bmp} \big] \,,
\end{eqnarray} 
where the $f_i$ are functions of $\bmp, \bmp^\prime$ and $\bmq$ that will be
given in Sec.\,\ref{sectionQCDm0m1}.  These operators are denoted graphically by
the diagram in Fig.~\ref{fig_us1}b, since they are essentially like the product
of two potentials summed over the intermediate 3-momentum ${\bf q}$. A similar
set of operators exists for three-loop diagrams with an ultrasoft gluon at order
$\alpha_s^4 v^{-1}$. To evaluate matrix elements of these operators using
dimensional regularization we combine the sums with the integration over the
ultrasoft spacetime coordinate $x$~\cite{LMR}, which leads to
\begin{eqnarray}
 \sum_{\bf q} \to \int d^{d-1}q  \,.
\end{eqnarray} 
Unlike the sums over labels on the fields, the free sum over ${\bf q}$ scales as
$v^{-2\epsilon}$, and contributes in determining the factor of $\mu_S^{4\epsilon}$
in Eq.~(\ref{Ok1a}).

At the level of Fig.~\ref{fig_us1}a it still appears ambiguous whether the sum
over ${\bf q}$ in Eq.~(\ref{Ok1a}) needs to be carried out. The key point is
that the original ultrasoft divergent loop in Fig.~\ref{fig_us1}a acts like a
one-loop subdivergence, despite the fact that it shows up at the level of the
two loop graph. Therefore, it renormalizes each term in the sum of the operators
in Eq.~(\ref{Ok1a}), rather than a potential $V({\bf p,p'})$ with the sum over
${\bf q}$ already carried out.

This renormalization prescription can be illustrated by considering a graph with
an additional ultrasoft fermion bubble as in Fig.~\ref{fig_us1}c.  To
renormalize Fig.~\ref{fig_us1}c we require both the $1/\epsilon$ counterterm for
the fermion bubble as in Fig.~\ref{fig_us1}d, and a $1/\epsilon^2$ counterterm
from Fig.~\ref{fig_us1}e.  which has an identical momentum structure to the
operators in Fig.~\ref{fig_us1}b.  In dimensional regularization the complete
result for the sum of the three graphs has the divergent structure
\begin{eqnarray}
  \Big(\frac{\mu_S^{2}}{{\bf k}^2}\Big)^\epsilon \bigg[ \frac{1}{2\epsilon^2}
    \Big(\frac{\mu_U^{2}}{E^2}\Big)^{2\epsilon} 
   - \frac{1}{\epsilon^2} \Big(\frac{\mu_U^{2}}{E^2}\Big)^{\epsilon} 
   + \frac{1}{2\epsilon^2} \bigg]  \,,
\end{eqnarray}
where in the limit $\epsilon\to 0$ all possible subdivergences
$\ln(E^2/\mu_U^2)/\epsilon$ and $\ln({\bf k}^2/\mu_S^2)/\epsilon$ are canceled
by the counterterm contributions. If we only needed counterterms to be added
after the sum on ${\bf q}$ was carried out then the sum of Figs.~\ref{fig_us1}c
and \ref{fig_us1}d
\begin{eqnarray} \label{bad}
  \Big(\frac{\mu_S^{2}}{{\bf k}^2}\Big)^\epsilon \bigg[ \frac{1}{2\epsilon^2}
    \Big(\frac{\mu_U^{2}}{E^2}\Big)^{2\epsilon} 
   - \frac{1}{\epsilon^2} \Big(\frac{\mu_U^{2}}{E^2}\Big)^{\epsilon} 
    \bigg] &=& -\frac{1}{2\epsilon^2} 
    - \frac{1}{2\epsilon} \ln\Big(\frac{\mu_S^2}{{\bf k}^2}\Big) 
    + \ldots \,,
\end{eqnarray}
should only contain an overall analytic divergence. However, Eq.~(\ref{bad}) has
a non-analytic divergence.  We note that as for the operators ${\cal
O}^{(\sigma)}_{2i}$ the Wilson coefficients ${\cal V}_{ki}(\nu)$ of the
operators ${\cal O}_{ki}$ vanish at the hard scale, i.e.  ${\cal
V}_{ki}(\nu=1)=0$. However, divergences from graphs such as those in
Fig.~\ref{fig_us1} lead to a non-zero anomalous dimension so that ${\cal
V}_{ki}(\nu<1)\neq 0$. Furthermore we emphasize that the renormalization of
these operators is not equivalent to the renormalization of the QCD $1/|{\bf
k}|$ and $1/{\bf k}^2$ potentials or to the renormalization of $1/r$ or $1/r^2$
potentials. This will be discussed further in Sec.~\ref{sectionQCDm0m1}.

\section{Muonic-Hydrogen with massless electrons} 
\label{sectionmH}

Our first example is a simplified toy model for muonic-Hydrogen with proton mass
$m_p\to\infty$, muon mass $m$ fixed, and $n_f$ massless ``electrons''.  In this
theory the running of the coupling is simplified because there are no gauge
boson self interactions. Furthermore at the order we are working the soft graphs
simply involve massless leptons. This model was proposed in
Refs.\,\cite{Pineda1,PinedaPC} as a test of the anomalous dimensions of the
$1/m^2$ operators (see also Ref.~\cite{Pineda3} and reference one in
\cite{Reviews}). In this section we show that vNRQCD correctly reproduces the
$n_f \alpha^6 \ln^2\alpha$ energy levels in this toy model and is therefore in
agreement with the QED limit of Pachuki's results in
Ref.~\cite{Pachucki}.

For this toy model we have $(T^A\otimes \bar T^A)\to (-1)$, $(1\otimes 1)\to
(+1)$ and therefore define ${\cal V}_i = {\cal V}_i^{(1)}-{\cal V}_i^{(T)}$. For
the soft spin-independent vertices with $m_p=\infty$ we only require
$Z_{0}^{(0)}=1/{\bf k}^2$ and $Z_0^{(2)}= -c_D(m\nu)/(8m^2)$ from
Ref.~\cite{amis}. Here $c_D(\mu)$ is the QED limit of the gauge invariant Darwin
coefficient computed in Ref.~\cite{BM},
\begin{eqnarray} \label{cDqed}
 c_D(\mu) = 1 + \frac{16}{3\bar\beta_0} \ln\bigg[ \frac{\alpha(\mu)}{\alpha(m)}
 \bigg] \,,
\end{eqnarray}
where $\bar\beta_0=-4n_f/3$.  The operator ${\cal O}_{2\varphi}^{(0)}$ has the
form
\begin{eqnarray}
{\cal O}_{2\varphi}^{(0)} 
& = &
{ e^4\,\mu_S^{4\epsilon}}\:  \:
(\psi_{\bmp^\prime}^\dagger\:\Gamma^{(0)}_\psi\: \psi_{\bmp}) \:
(\chi_{-\bmp^\prime}^\dagger\:\Gamma^{(0)}_\chi\: \chi_{-\bmp})
   \ (\overline\varphi_{-q}\: \Gamma_ \varphi^{(0)}\: 
  \varphi_{q}) \
\,, 
\label{OdefQED}
\end{eqnarray}
where
\begin{eqnarray}
\Gamma_\varphi^{(0)}(q,\bmp,\bmp^\prime)
& = &
\bigg[\,
\frac{(2q^0\gamma^0+{\bf k}\cdot{\mbox{\boldmath$\gamma$}})}
 {{\bf k}^2+2 {\bf k}\cdot {\bf q}} +
 \frac{(2q^0\gamma^0-{\bf k}\cdot{\mbox{\boldmath $\gamma$}})}
 {{\bf k}^2-2 {\bf k}\cdot {\bf q}}
\,\bigg]\,\Big(Z_0^{(0)}\Big)^2
\,,
\nonumber\\[2mm]
\Gamma^{(0)}_{\varphi,\psi} & = & 1\,,\quad \Gamma^{(0)}_{\varphi,\chi}
\, = \, -1 \,,
\end{eqnarray}
and $\bmk=\bmp^\prime-\bmp$.  From tree level matching we have
$C_{2\varphi}^{(0)}(1)=0$.  Furthermore, since ultrasoft photons bring an extra
$v^2$, there is no anomalous dimension induced for this coefficient, so in fact
$C_{2\varphi}^{(0)}(\nu)=0$ and this operator does not need to be considered
further. The operator ${\cal O}_{2\varphi}^{(2)}$ has the form
\begin{eqnarray}
{\cal O}_{2\varphi}^{(2)}
& = &
-\frac{\bmk^2}{6\,m^2}\,{\cal O}_{2\varphi}^{(0)}
\,.
\end{eqnarray}
From tree level matching one finds that its coefficient vanishes also at the
hard scale, $C_{2\varphi}^{(2)}(1)=0$, but it gets a non-trivial anomalous
dimension from the UV divergences in the ultrasoft graph in Fig.~\ref{fig_sus}b
plus wavefunction counterterms,
\begin{eqnarray} \label{F2}
 \begin{picture}(60,40)(1,20)
  \includegraphics[width=0.8in]{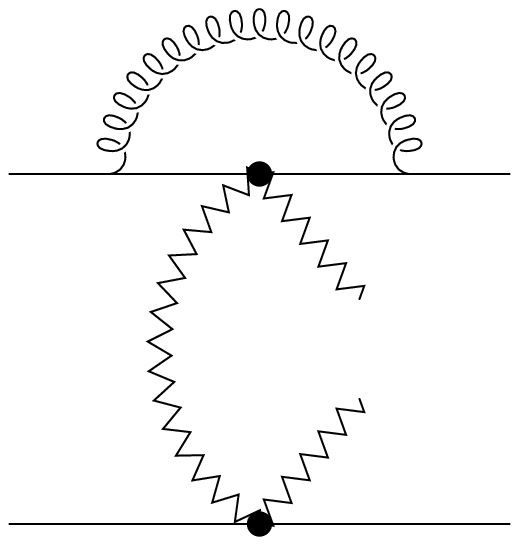} 
 \end{picture} \mbox{+ w.fn.}
& = & 
 -\frac{1}{6\,\pi}\,\frac{\bmk^2}{m^2}\,
  \frac{\alpha(m\nu^2)}{\epsilon}\,
  \langle i{\cal O}_{2\varphi}^{(0)}\rangle_{4e2\gamma}
\, = \,
 \frac{1}{\pi}\,
  \frac{\alpha(m\nu^2)}{\epsilon}\,
  \langle i{\cal O}_{2\varphi}^{(2)}\rangle_{4e2\gamma}
\,, \\[0pt]\nn
\end{eqnarray}
where the soft vertices symbolized by the dots stand for the coupling
$Z_0^{(0)}$. Together with the respective soft divergence induced by the pull-up
mechanism~\cite{hms1} this gives
\begin{eqnarray}
& & 
 \frac{1}{\pi}\, \bigg[\,
  \frac{\alpha(m\nu^2)}{\epsilon} - \frac{\alpha(m\nu)}{\epsilon} \,\bigg]\, 
  \langle i{\cal O}_{2\varphi}^{(2)}\rangle_{4e2\gamma} \,. 
\label{rslt}
\end{eqnarray}
We see that the result would vanish if the two couplings were
evaluated at the same scale. The result in Eq.~(\ref{rslt}) is canceled by a
counterterm for $C_{2\varphi}^{(2)}$ which, using the vRGE, gives the anomalous
dimension
\begin{eqnarray}
 \nu\frac{\partial}{\partial\nu} C_{2\varphi}^{(2)} = 
  -\frac{2}{\pi} \Big[\, 2\alpha(m\nu^2) - \alpha(m\nu) \,\Big] \,.
\label{vRGEQED}
\end{eqnarray}
There is another possible contribution to the anomalous dimension in
Eq.\,(\ref{vRGEQED}) coming from the soft  coupling renormalization graphs
for the operator ${\cal O}^{(2)}_{2\varphi}$, however these are exactly canceled
by the charge counterterm contribution associated to the factor $e^4$ in
Eq.\,(\ref{OdefQED}). Similarly, an UV divergence in the analogue of
Fig.\,\ref{fig_sus}b having only soft gluons is exactly canceled by a
counterterm associated to the $Z^{(2)}_0$ term in the soft Lagrangian.  Solving
Eq.~(\ref{vRGEQED}) with the boundary condition $C_{2\varphi}^{(2)}(1)=0$ gives
\begin{eqnarray} \label{C2sln}
 C_{2\varphi}^{(2)}(\nu) 
& = &
 \frac{4}{\beta_0}\, \ln\bigg[ \frac{\alpha(m\nu^2)}{\alpha(m\nu)} \bigg]
 \, = \, \frac{3}{4}\,\Big[\,c_D(m\nu^2) - c_D(m\nu)\,\Big] \,.
\end{eqnarray}

\begin{figure}[t!]
\centerline{ 
\raisebox{0.cm}{\includegraphics[width=1.4in]{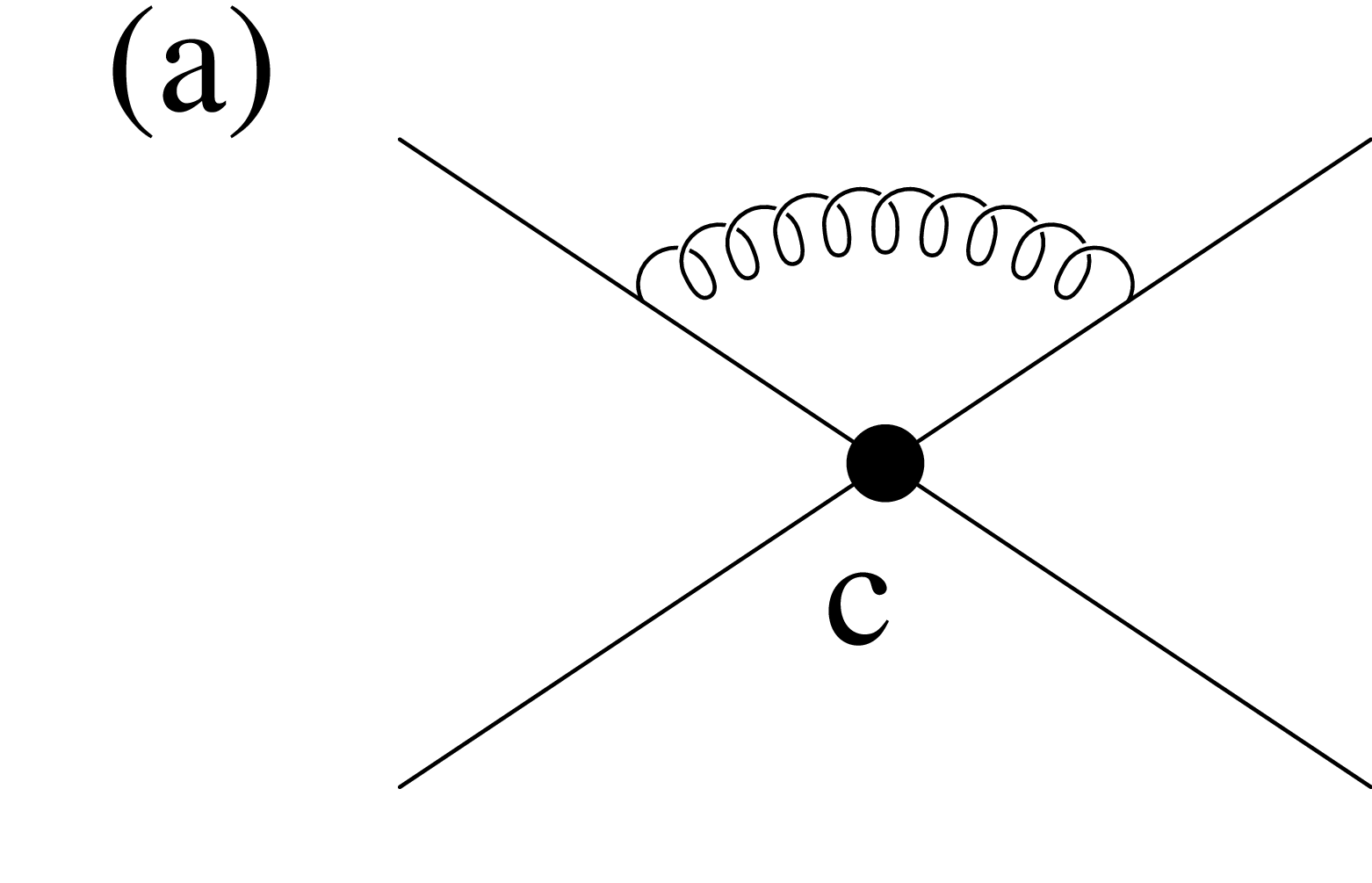}}
\hspace{0.5cm}
\raisebox{0.15cm}{\includegraphics[width=1.in]{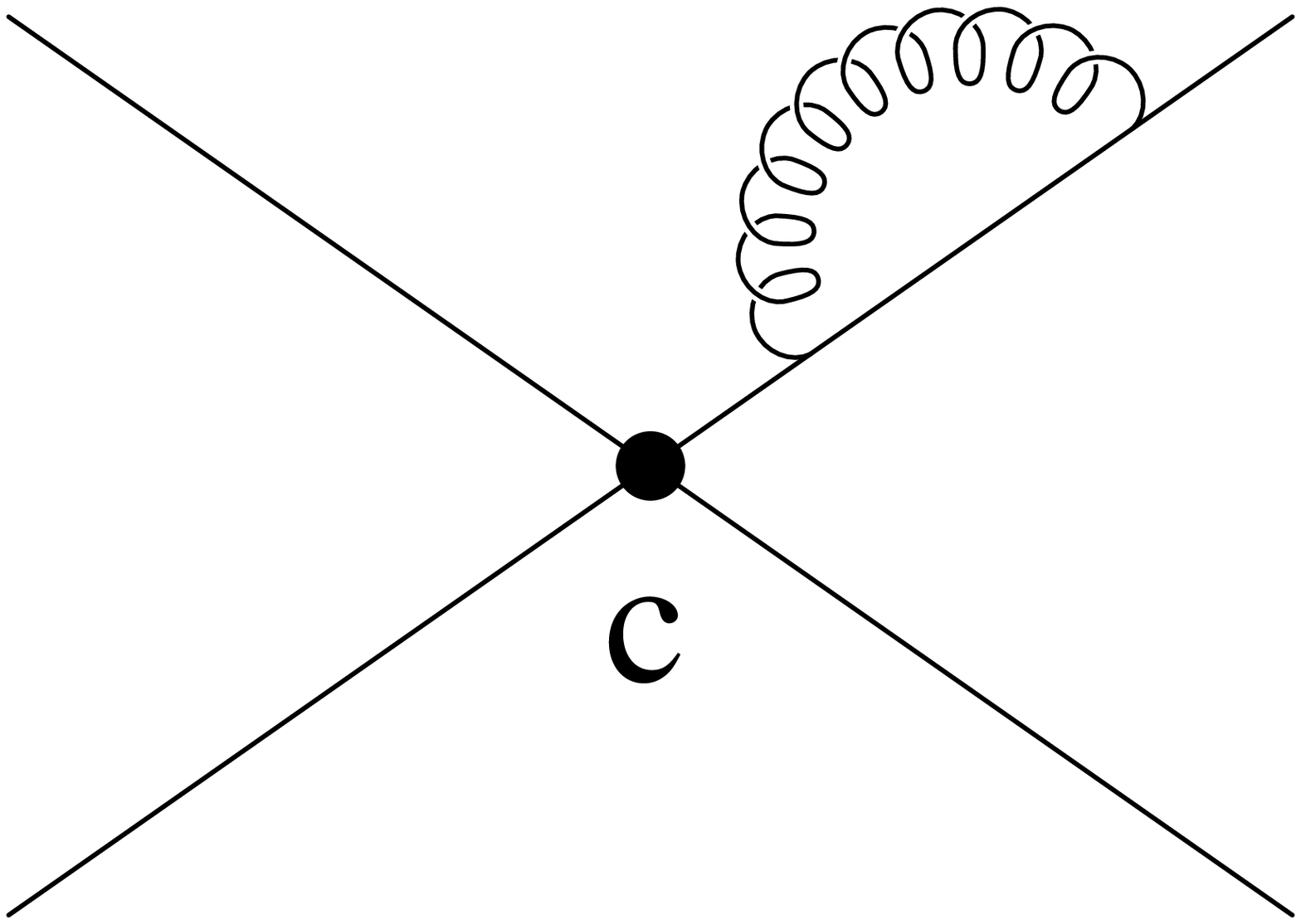} } 
\hspace{0.7cm}
\raisebox{-0.5cm}{\includegraphics[width=1.3in]{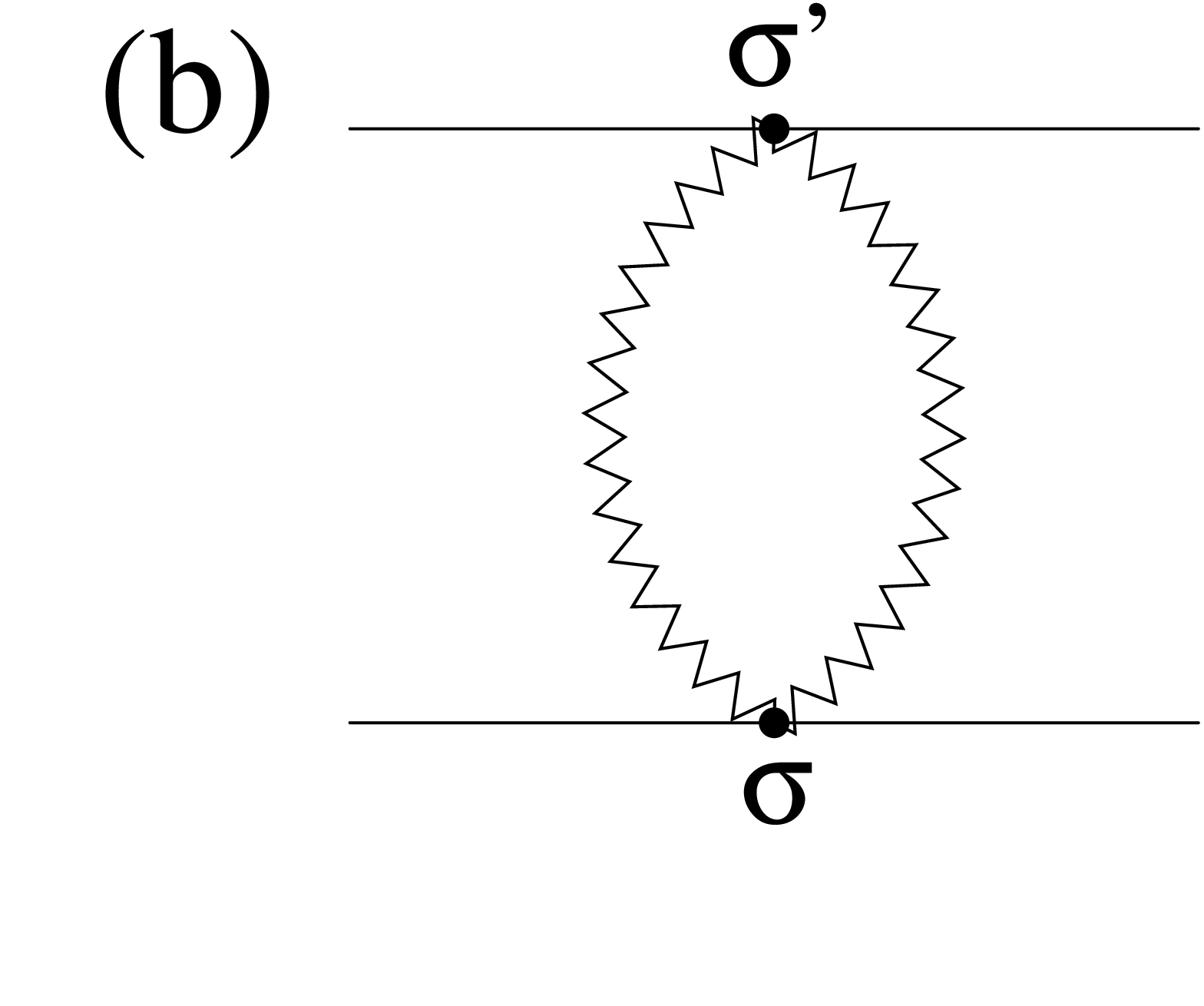} }
\hspace{0.7cm}
\includegraphics[width=1.3in]{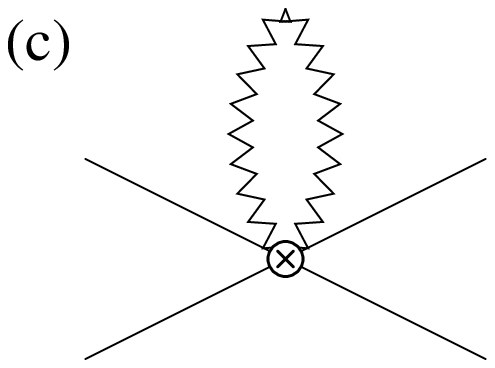}
 }
\vspace{-0.5cm}
\caption{a) ultrasoft gluon graphs with ${\bf p}\cdot {\bf A}$ vertices (with
wavefunction renormalization on the other line understood), b) soft graph with
$Z_0^{(\sigma=0)}$ and $Z_0^{(\sigma'=2)}$ vertices, c) soft graph involving
${\cal O}_{2\varphi}^{(2)}$. The zig-zag lines here denote massless soft
fermions.
\label{fig_s3}}
\end{figure}
The anomalous dimension for ${\cal V}_2$ can now be computed from the graphs in
Fig.~\ref{fig_s3} which are order $v^1 e^4$. The graphs in Fig.~\ref{fig_s3}a,
\ref{fig_s3}b,
\ref{fig_s3}c give the following three terms
\begin{eqnarray}
 \nu \frac{\partial}{\partial\nu} {\cal V}_2 &=& -\frac{8}{3} \:
 \alpha(m\nu)\: \alpha(m\nu^2)+ \frac{n_f}{3}\: c_D(m\nu)\:[\alpha(m\nu)]^2
 + \frac{4 n_f}{9}\: C_{2\varphi}^{(2)}(\nu)\: [\alpha(m\nu)]^2 \,.
\end{eqnarray}
Using the solution in Eq.~(\ref{C2sln}) this becomes
\begin{eqnarray}
 \nu \frac{\partial}{\partial\nu} {\cal V}_2 
 &=& -\frac{8}{3} \: \alpha(m\nu)\: \alpha(m\nu^2)
    +\frac{n_f}{3}\: c_D(m\nu^2)\:[\alpha(m\nu)]^2 \,.
\end{eqnarray}
Solving this equation with the boundary condition ${\cal V}_2(1)=\pi\alpha(m)/2$
we find
\begin{eqnarray}
{\cal V}_2(\nu) =  
 \frac{\pi}{2}\: c_D(m\nu^2)\: \alpha(m\nu) \,.
\end{eqnarray}
The result differs from the coefficient of the four quark operator in
Ref.~\cite{Pineda3} in \mNRQCD ($mv<\mu<m$). Our result is in agreement with the
pNRQCD result in Ref.~\cite{Pineda3}
\begin{eqnarray}
 \pi\, D_d^{(2)}(\nu_{us})= \frac{\pi}{2}\: c_D(\nu_{us})\: \alpha(r^{-1})\,,
\end{eqnarray}
but only if we correlate the \mpNRQCD matching scale $1/r$ with the pNRQCD
renormalization scale $\nu_{us}$ for the energy by enforcing $\nu_{us}=m\nu^2$
and $1/r=m\nu$.  We also agree with Pachucki~\cite{Pachucki} for the
$n_f\,\alpha^6\ln^2\alpha$ energy levels for this toy model. In
Ref.~\cite{Pineda3} a different expression for ${\cal V}_2(\nu)$ was inferred
for the vNRQCD approach, however this is because the contribution from the graph
in Fig.~\ref{fig_s3}c was missing.

\section{QCD results for $1/m^2$ potentials}
\label{sectionQCDm2}

Our second example of the effect of the operators in Eq.~(\ref{6q}) is
non-relativistic QCD for equal mass heavy fermions. In this case the soft
degrees of freedom include soft gluons, quarks, and ghosts. The operators ${\cal
O}_{2\varphi}^{(0)}$, ${\cal O}_{2A}^{(0)}$ and ${\cal O}_{2c}^{(0)}$ can be
written in a compact form in terms of the soft vertices $U_{\mu\nu}^{(0)}$,
$Y^{(0)}$, and $Z^{(0)}$ from Ref.~\cite{amis}, as summarized in Appendix
A. Closing the respective two soft (gluon, quark and ghost) lines one obtains
for the one-loop four quark matrix element a structure that is identical to the
one-loop time-ordered product of the soft vertices $U_{\mu\nu}^{(0)}$,
$Y^{(0)}$, and $Z^{(0)}$ [up to non-trivial Wilson coefficients which are
suppressed in this equality]:
\begin{eqnarray}
 \begin{picture}(70,35)(1,40)
  \put(-15,25){
  \includegraphics[width=1.1in]{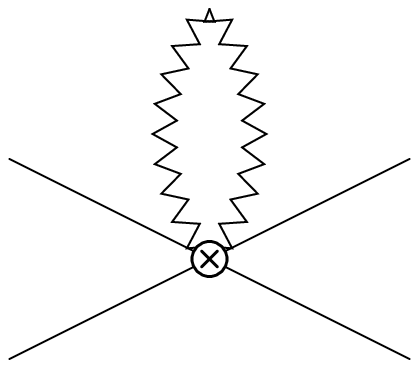}}
 \end{picture} 
 &=& 
\langle i{\cal O}_{2A}^{(0)} \rangle_{4Q}
\, + \,
\langle i{\cal O}_{2c}^{(0)} \rangle_{4Q}
\, + \,
\langle i{\cal O}_{2\varphi}^{(0)} \rangle_{4Q}
 = \sum_i \langle i{\cal O}_{2i}^{(0)} \rangle_{4Q}
\nonumber\\[2mm]
& = &
-i\: \frac{\beta_0\,\alpha_s(m\nu)^2}{{\bf k}^2\,\epsilon} 
\: T^A\otimes\bar T^A  +\ldots 
= \begin{picture}(70,30)(1,40)
  \put(0,0){
  \includegraphics[width=0.8in]{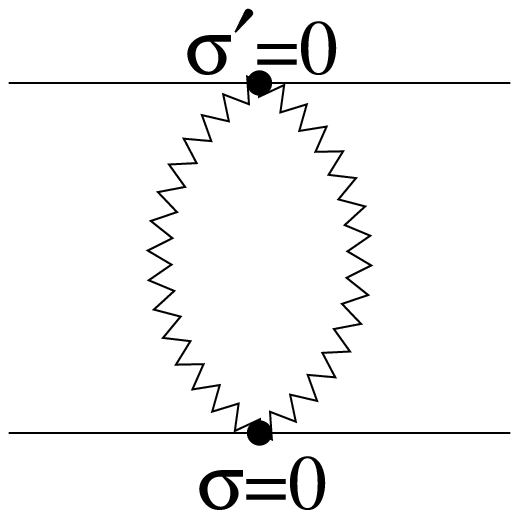}}
 \end{picture} 
\,. \\[5pt]\nn
\end{eqnarray}
Just as in the matching calculation for the QED toy model the momentum structure
of the soft time-ordered products agrees with the full theory result, so at
$\nu=1$ we find $C_{2i}^{(0)}(1)=0$. Similarly, since ultrasoft gluons bring an
extra factor $v^2$ there is again no anomalous dimension induced for the
operators ${\cal O}_{2i}^{(0)}$ and we have $C_{2i}^{(0)}(\nu)=0$ identically.

For $\sigma=2$ it is necessary to keep the different color structures between
the heavy fermions, so in general we have to consider two types of operators,
${\cal O}_{2i}^{(2),(T)}$ with the fermionic structure
$[\psi_{\bmp^\prime}^\dagger T^A \psi_{\bmp} \chi_{-\bmp^\prime}^\dagger \bar
T^A \chi_{-\bmp}]$ and ${\cal O}_{2i}^{(2),(1)}$ with the fermionic structure
$[\psi_{\bmp^\prime}^\dagger \psi_{\bmp}\chi_{-\bmp^\prime}^\dagger
\chi_{-\bmp}]$. These operators are defined by Eq.~(\ref{O2abc}) with ${\cal
O}_{2i}^{(0),(1,T)}$ defined in Appendix A. For our purposes it is sufficient to
consider only operators where the soft lines are closed in color space, and the
operators ${\cal O}_{2i}^{(0)(1,T)}$ and ${\cal O}_{2i}^{(2)(1,T)}$ are defined
with this convention.  This is sufficient because for the renormalization of the
potentials all external soft lines are contracted.  From tree level matching one
obtains that the coefficients of these operators vanish at the hard scale
$C_{2i}^{(2),(T)}(1)=C_{2i}^{(2),(1)}(1)=0$, but as in the QED case they have a
non-vanishing anomalous dimension due to the exchange of ultrasoft gluons in
diagrams of order $\alpha_s^2 v^1$ as shown in Fig.~\ref{fig_sus}b.  Now since
we are dealing with equal mass fermions, the ${\bf p}\cdot {\bf A}$ ultrasoft
gluons can attach to any of the heavy fermion lines.  Including the permutations
of possible ultrasoft attachments we find
\begin{eqnarray} \label{F2qcd}
  \begin{picture}(140,40)(1,20)
  \includegraphics[width=0.8in]{figs/softQ3}  \!\raisebox{0.7cm}{+}
  \includegraphics[width=0.8in]{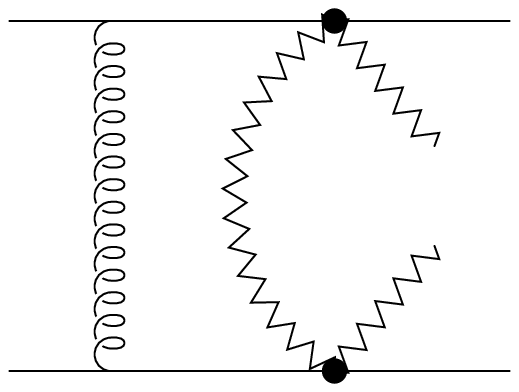} 
 \end{picture} \!\!\!\!\mbox{+ $\ldots$ }
&=& 
\frac{\alpha_s(m\nu^2)}{\pi\epsilon}\, \sum_i
\bigg\{
\frac{C_1}{3}\,\frac{\bmk^2}{m^2}\,
\langle i{\cal O}_{2i}^{(0),(1)} \rangle_{4Q2i}
-\!
\bigg[\, C_A\,\frac{(\bmp^2+\bmp^{\prime\,2})}{3m^2} \nn\\
&& +
\Big(\frac{C_F}{3}+\frac{C_d}{12}-\frac{C_A}{4}\Big)\,\frac{\bmk^2}{m^2}
\,\bigg]\,\langle i{\cal O}_{2i}^{(0),(T)} \rangle_{4Q2i}
\,\bigg\}\,,
\end{eqnarray}
where the ellipses denote other attachments including wavefunction diagrams.
For SU($N_c$) QCD the color coefficients that appear here are 
\begin{eqnarray}
 C_A=N_c \,,\qquad C_F=\frac{N_c^2-1}{2N_c} \,,\qquad 
 C_d=8C_F-3C_A \,, \qquad C_1=\frac{1}{2}\,C_F C_A-C_F^2\,. 
\end{eqnarray}
The result in Eq.~(\ref{F2qcd}) renormalizes the operators
\begin{eqnarray} \label{O2abc}
 {\cal O}_{2a}^{(2),(1)} &=&  \frac{\bmk^2}{m^2}\, \sum_i {\cal O}_{2i}^{(0),(1)}
   \,,\quad
 {\cal O}_{2b}^{(2),(T)} =  \frac{\bmk^2}{m^2}\, \sum_i {\cal O}_{2i}^{(0),(T)}
   \,,\quad \nn\\
 {\cal O}_{2c}^{(2),(T)} &=&  \frac{(\bmp^2+\bmp^{\prime\,2})}{m^2}\, 
   \sum_i {\cal O}_{2i}^{(0),(T)} \,,
\end{eqnarray}
whose coefficients $C_{2a}^{(2)}$, $C_{2b}^{(2)}$, and $C_{2c}^{(2)}$ vanish at
the matching scale $\nu=1$. Adding to Eq.~(\ref{F2qcd}) the soft divergences
induced by the pull-up mechanism one arrives at the following anomalous
dimension for the coefficients:
\begin{eqnarray}
 \nu\frac{\partial}{\partial\nu} C_{2a}^{(2)} 
   &=&   \frac{-2C_1}{3\pi}\gamma_{u2}(\nu) \,,\quad
 \nu\frac{\partial}{\partial\nu} C_{2b}^{(2)} 
   = \frac{4C_F\!+\!C_d\!-\!3C_A}{6\pi} \gamma_{u2}(\nu) \,,\quad
 \nu\frac{\partial}{\partial\nu} C_{2c}^{(2)}
   =  \frac{2C_A}{3\pi} \gamma_{u2}(\nu) \,,\nn\\[5pt]
 \gamma_{u2}(\nu) &\equiv& 2\alpha_s(m\nu^2) - \alpha_s(m\nu) \,.
\label{vRGEQCD}
\end{eqnarray}
These equations are quite similar to Eq.\,(\ref{vRGEQED}) and their solutions
are
\begin{eqnarray} \label{C2slnQCD}
 C_{2a}^{(2)}(\nu) & = & \frac{4C_1}{3\beta_0}
    \ln(w)
  \,,\qquad\quad
 C_{2b}^{(2)}(\nu) =  \frac{3C_A\!-\!C_d\!-\!4C_F}{3\beta_0}
    \ln(w)
   \,,\nn\\[2mm]
 C_{2c}^{(2)}(\nu) &=& \frac{-4C_A}{3\beta_0}\, 
  \ln(w)
 \,,
\label{solutionvRGEQCD}
\end{eqnarray}
where
\begin{eqnarray} \label{w}
  w = \frac{\alpha_s(m\nu^2)}{\alpha_s(m\nu)}  \,.
\end{eqnarray} 
Note that as in the QED case there are additional possible contributions to the
anomalous dimensions of the operators ${\cal O}_{2i}^{(2),(1,T)}$ which are,
however, exactly canceled by pre-existing counterterms. This includes for
example the purely soft diagram analogous to Fig.\,\ref{fig_sus}b, and the soft
coupling renormalization graphs. The divergences of the soft diagrams are
canceled by counterterms for $U_{\mu\nu}^{(2)}$, $Z_0^{(2)}$, $Y^{(2)}$ and the
divergences of the soft coupling graphs by the counterterm associated to the
factor $g_s^4$ in Eq.~(\ref{6q}).

\begin{figure}[t!]
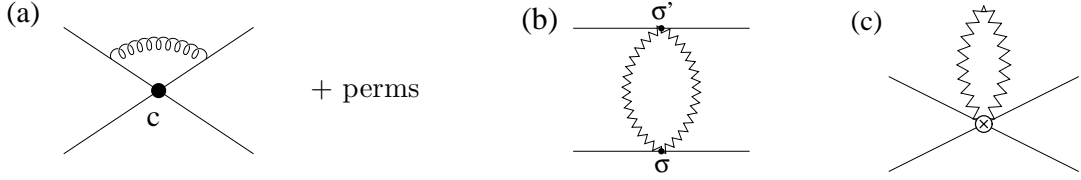

\centerline{ 
\raisebox{0.cm}{\includegraphics[width=1.4in]{figs/usoft2}}
\hspace{0.5cm}
\raisebox{1cm}{\mbox{+ perms} } 
\hspace{0.7cm}
\raisebox{-0.5cm}{\includegraphics[width=1.3in]{figs/9923_fd8} }
\hspace{0.7cm}
\includegraphics[width=1.3in]{figs/soft3}
 }
\vspace{-0.5cm}
\caption{a) ultrasoft gluon graphs with ${\bf p}\cdot {\bf A}$ vertices on all
permutations of the four fermion lines, b) soft graph with
$Z_0^{(\sigma,\sigma')}$, $U_{\mu\nu}^{(\sigma,\sigma')}$ or
$Y^{(\sigma,\sigma')}$ vertices such that $\sigma+\sigma'=2$, c) soft graph
involving one of the ${\cal O}_{2a}^{(2)}$, ${\cal O}_{2b}^{(2)}$, ${\cal
O}_{2c}^{(2)}$ operators.
\label{fig_s4}}
\end{figure}

Finally, we consider the anomalous dimensions for the spin-independent potential
coefficients ${\cal V}_{2,r}^{(1,T)}(\nu)$. The graphs required for this
computation are shown in Fig.~\ref{fig_s4}. The graphs in Fig.~\ref{fig_s4}a and
\ref{fig_s4}b were computed in Ref.~\cite{amis}, however the contributions
from the graph in Fig.~\ref{fig_s4}c which are proportional to the coefficients
$C_{2i}^{(2)}$ were not included.
Adding the contributions from all three graphs the results are 
\begin{eqnarray}  \label{VRG}
\nu \frac{\partial}{\partial\nu} {\cal V}_r^{(T)} &=& -2 (\beta_0 + \frac{8}{3}
  C_A)\, \alpha_s^2(m\nu) + \frac{32}{3} C_A\, \alpha_s(m\nu) \alpha_s(m\nu^2)
  -4\,\beta_0\, C_{2c}^{(2)}(\nu)\, \alpha_s(m\nu)^2 \,, 
\nn\\[3pt]
\nu \frac{\partial}{\partial\nu} {\cal V}_2^{(T)} &=&
  \left\{ \frac{\beta_0}2 \Big[1+c_D(\nu)-2 c_F^2(\nu) \Big]+
  \frac{C_A}{6}  \Big[28 - 11 c_D(\nu) + 26 c_F(\nu)^2 \Big] -
  \frac{7C_d}{6}  \right\} \, \alpha_s^2(m\nu) 
     \nn\\[3pt] &&  
+ \,\frac{4}{3} (4C_F+C_d-3C_A)\, \alpha_s(m\nu) \alpha_s(m\nu^2) 
 -2\,\beta_0\, C_{2b}^{(2)}(\nu)\, \alpha_s(m\nu)^2 
  \,,  \nn\\[3pt]
\nu \frac{\partial}{\partial\nu} {\cal V}_2^{(1)} &=& \frac{14}{3}\, C_1\,
  \alpha_s^2(m\nu) - \frac{16}{3}\,C_1\, \alpha_s(m\nu) \alpha_s(m\nu^2)
  -2\,\beta_0\, C_{2a}^{(2)}(\nu)\, \alpha_s(m\nu)^2  \,.
\end{eqnarray}
The gauge invariant HQET coefficients that appear here are~\cite{BM}
\begin{eqnarray}
  c_F(\nu) = z^{-C_A/\beta_0}\,, \quad
  c_D(\nu) = z^{-2C_A/\beta_0} +\Big(\frac{20}{13}+\frac{32 C_F}{13 C_A}\Big)
  \big[ 1 - z^{-13C_A/(6\beta_0)} \big]  \,,
\end{eqnarray}
where $z=\alpha_s(m\nu)/\alpha_s(m)$. The solutions for the coefficients in
Eq.~(\ref{VRG}) are
\begin{eqnarray}  \label{pv2}
 {\cal V}_r^{(T)}(\nu) &=& 4\pi\,\alpha_s(m)\,z  
  - \frac{32\pi C_A}{3\beta_0}\, \alpha_s(m)\,z\,\ln(w) \,, \\[5pt]
 {\cal V}_2^{(T)}(\nu) &=& \frac{ \pi \left[ C_A (48C_F + 13 C_d +4C_A) 
   -\beta_0(33C_A + 32 C_F) \right]}
   { 13\beta_0\, C_A }\,\alpha_s(m)\,\left(z - 1\right) \nn\\[3pt] 
 & + & \frac{8 \pi(3\beta_0 - 11C_A)(5C_A + 8C_F)}
  {13 C_A(6\beta_0 - 13 C_A)}\,\alpha_s(m)\,
  \left[z^{1 - 13C_A/(6\beta_0)} - 1\right] \nn\\[3pt] 
 & + & \frac{\pi(\beta_0 - 5C_A)}{(\beta_0 - 2C_A)}\,\alpha_s(m)\,
 \left[z^{1 - 2C_A/\beta_0} - 1\right] 
 -\frac{4\pi(4C_F + C_d - 3C_A)}{3\beta_0}\,\alpha_s(m)\,z \ln(w) 
 \,, \nn \\[5pt]
 {\cal V}_2^{(1)}(\nu) &=& \frac{4\pi C_1}{\beta_0}\,\alpha_s(m)\,
  \left(1 - z\right) +\frac{16\pi C_1}{3\beta_0}\,\alpha_s(m)\,z\ln(w)
  \,. \nn 
\end{eqnarray}
For the color singlet channel ${\cal V}_i^{(s)}={\cal V}_i^{(1)}
- C_F {\cal V}_i^{(T)}$ and the above results give
\begin{eqnarray} \label{Vsrslts}
 {\cal V}_r^{(s)}(\nu) &=& -4\pi\,C_F\,\alpha_s(m)\,z\, \Big[ 1 - 
  \frac{8 C_A}{3\beta_0}\,\ln(w) \Big]\,, \\[5pt]
 {\cal V}_2^{(s)}(\nu) &=& \pi C_F \,\alpha_s(m)\,\left(z - 1\right) 
 \left[ \frac{33}{13} +\frac{32 C_F}{13 C_A}+ \frac{9C_A}{13\beta_0} 
 -\frac{100C_F}{13\beta_0}\right] \nn \\[3pt]
 &-& \frac{8 \pi C_F(3\beta_0\! -\! 11C_A)(5C_A \!+\! 8C_F)}
  {13\, C_A(6\beta_0\! -\! 13 C_A)}\,\alpha_s(m)\,
  \left[z^{1 - (13C_A)/(6\beta_0)} - 1\right] \nn\\[3pt] 
 &-& \frac{\pi\, C_F (\beta_0 \!-\! 5C_A)}{(\beta_0 \!-\! 2C_A)}\,\alpha_s(m)\,
 \left[z^{1 - 2C_A/\beta_0} - 1\right] 
 -\frac{16\pi C_F(C_A\!-\! 2C_F)}{3\beta_0}\,\alpha_s(m)\,z \ln(w) \,. \nn 
\end{eqnarray} 
The results in Eqs.~(\ref{pv2}) and (\ref{Vsrslts}) are valid for all values of
$\nu$ between $1$ and a $v_0$ of order the physical velocity of the quarks. This
accounts for energy scales between $m$ and $mv^2$ and momentum scales between
$m$ and $mv$, so our results apply to this {\em entire} region. Our results do
not agree with the running of the 4-quark operators $d_{ss}$, $d_{sv}$,
$d_{vs}$, $d_{vv}$ obtained in Ref.~\cite{Pineda1} for $mv < \mu < m$
(\mNRQCD). This is because in Ref.~\cite{Pineda1} ultrasoft gluons are not
included in the results for $\mu > mv$. To translate the results in
Ref.~\cite{Pineda1} for scales $\mu < mv$ we note that ${\cal V}_r^{(s)}=-4\pi
C_F\, D_{1,s}^{(2)}$, ${\cal V}_2^{(s)} = \pi C_F
(D_{d,s}^{(2)}-D_{2,s}^{(2)})$, and we again enforced a correlation by demanding
that the pNRQCD renormalization scale $\nu_{us}=m\nu^2$, and the \mpNRQCD
matching scale $1/r=m\nu$.  With these restrictions our results in
Eq.~(\ref{Vsrslts}) agree with Ref.~\cite{Pineda1} for the case $\mu < mv$.

Note that there are also constraints on the relation between the cutoff for the
momentum ${\bf p}$ of the quarks and the cutoff for the momentum transfers ${\bf
k}={\bf p'}-{\bf p}$. Reproducing the known $\alpha^7\ln^2\alpha$ hyperfine
splitting for positronium~\cite{a7ln2}, requires that the cutoffs for these
scales are correlated, since the calculation depends on simultaneously
integrating anomalous dimensions that arise from soft and potential
loops~\cite{amis4}.

The remaining coefficients of the $1/m^2$ potentials are spin-dependent and not
affected by the operator in Fig.~\ref{fig_sus}c. They were computed in 
Ref.~\cite{amis}:
\begin{eqnarray} \label{Vspin}
 {\cal V}_s^{(T)}(\nu) &=& \frac{2 \pi}{(2 C_A-\beta_0) } \,
  \alpha_s(m) \, \left[
   C_A + \frac{1}{3} ( 2\beta_0 - 7 C_A) \ z^{(1-2 C_A/\beta_0)} \right]
    \,, \\[3pt]
%
%
 {\cal V}_t^{(T)}(\nu) &=& -\frac{\pi}{3}\,\alpha_s(m)\,
    \ z^{(1-2 C_A/\beta_0)}  \,,\nn \\[3pt]
 {\cal V}_\Lambda^{(T)}(\nu) &=& 2 \pi\,\alpha_s(m) \left[ z - 4\
    z^{(1-C_A/\beta_0)}  \right] \nn \,.
\end{eqnarray}
These expressions for ${\cal V}_t^{(T)}$ and ${\cal V}_{\Lambda}^{(T)}$ agree
with Ref.~\cite{chen} (however ${\cal V}_s^{(T)}$ disagrees).  The results for
the coefficients in Eq.~(\ref{Vspin}) were confirmed in Ref.~\cite{Pineda1}.


\section{QCD results for operators ${\cal O}_{ki}$, ${\cal O}_{ci}$ and 
$1/|{\bf k}|$, $1/{\bf k}^2$ potentials}
\label{sectionQCDm0m1}

\begin{figure}[t!]
\centerline{ 
\includegraphics[width=1.4in]{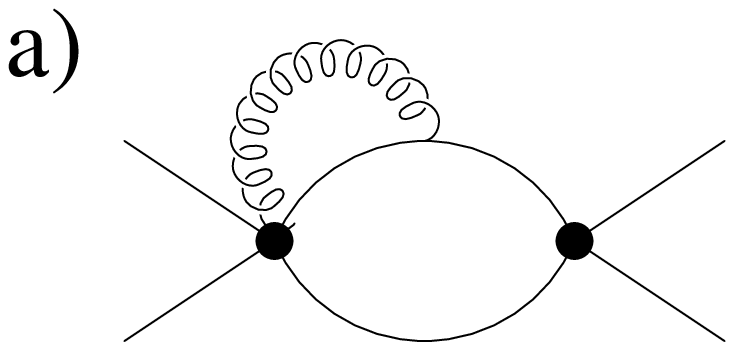} \hspace{0.2cm}
\includegraphics[width=1.4in]{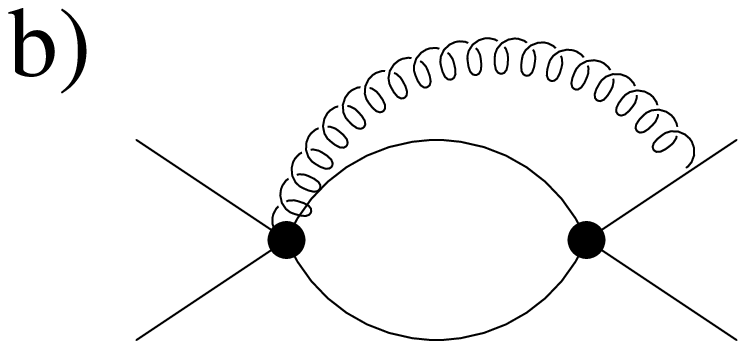} \hspace{0.3cm}
\includegraphics[width=1.4in]{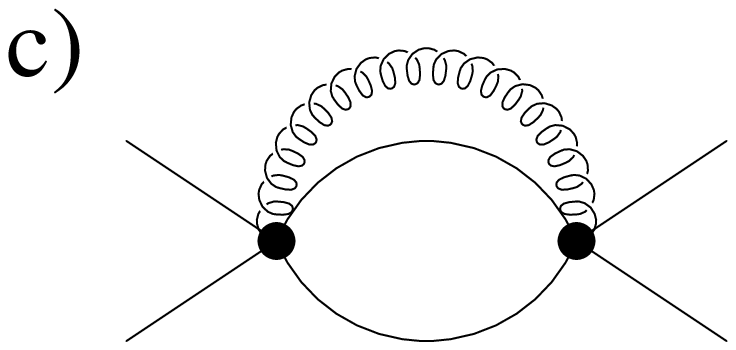} \hspace{0.3cm}
\includegraphics[width=1.3in]{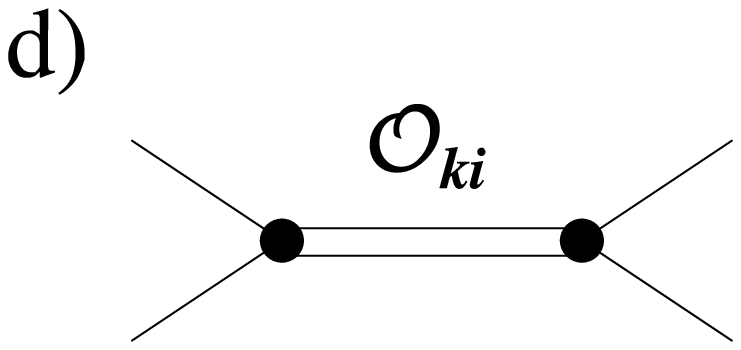} \hspace{0.2cm}  }
\vspace{-0cm} \caption{Ultrasoft graphs for the renormalization of the 
operators ${\cal O}_{k1}^{(1,T)}$ (left-right and up-down symmetric graphs
are implied).
\label{fig_usk}}
\end{figure}

At order $\alpha^3 v^0$ there are three diagrams which have ultrasoft
gluons (shown in Figs.~\ref{fig_usk}a,b,c) whose divergences are not
canceled by counterterms from ${\cal V}_{2,r}$~\cite{amis3}. As
pointed out in Sec.\,\ref{sect_usoft} these diagrams should be
renormalized by operators that do not have the sum over intermediate
potential quark 3-momenta ${\bf q}$ carried out, as pictured in 
Fig.~\ref{fig_usk}d.
Evaluating Figs.~\ref{fig_usk}a,b,c we find that the required
operators have the form
\begin{eqnarray} \label{Ok1}
{\cal O}_{k1}^{(1)} &=& - \frac{[\mu_S^{2\epsilon}\,{\cal V}_c^{(T)}]^2}{m}\: 
 \sum_{{\bf p,p',q}} ( f_{0} + f_{1} + 2 f_{2} )\
 \big[ \psi_{\bmp^\prime}^\dagger \psi_{\bmp}
 \chi_{-\bmp^\prime}^\dagger \chi_{-\bmp} \big] \,, \nn\\
{\cal O}_{k2}^{(T)} &=& - \frac{[\mu_S^{2\epsilon}\,{\cal V}_c^{(T)}]^2}{m}\: 
 \sum_{{\bf p,p',q}} ( f_{1} + f_{2} ) \ 
 \big[ \psi_{\bmp^\prime}^\dagger T^A \psi_{\bmp}
 \chi_{-\bmp^\prime}^\dagger \bar T^A \chi_{-\bmp} \big] \,,
\end{eqnarray}
giving the contribution $\Delta {\cal L}_p = {\cal V}_{k1}^{(1)} {\cal
O}_{k2}^{(1)} + {\cal V}_{k2}^{(T)} {\cal O}_{k1}^{(T)}$ to the vNRQCD
Lagrangian with Wilson coefficients ${\cal V}_{k1}^{(1)}$ and ${\cal
V}_{k2}^{(T)}$, respectively. The $\mu_S^\epsilon$ factors in Eq.~(\ref{Ok1})
are determined as in sections~\ref{sect_formalism},\ref{sect_usoft}. In
Eq.~(\ref{Ok1}) the functions $f_i$ are
\begin{eqnarray}
 && f_0 = \frac{\bmp^\prime\cdot (\bmq-\bmp)}{(\bmq-\bmp)^4\,
  (\bmq-\bmp^\prime)^2} 
  + (\bmp \leftrightarrow \bmp^\prime)\,, \qquad
 f_1 = \frac{\bmq \cdot (\bmq-\bmp)}{(\bmq-\bmp)^4\,(\bmq-\bmp^\prime)^2} 
  + (\bmp \leftrightarrow \bmp^\prime)\,, \nn\\
 && f_2 = \frac{(\bmq-\bmp^\prime)\cdot (\bmq-\bmp)}
  {(\bmq-\bmp)^4\,(\bmq-\bmp^\prime)^4}\: (\bmq^2-\bmp^{\prime\,2}/2-\bmp^2/2) 
  \,.
\end{eqnarray}
If the finite sums over ${\bf q}$ were carried out as integrals $d^3q$ or $d^nq$
with $n=d-1\to 3$ then the operators ${\cal O}_{ki}^{(1,T)}$ would reduce to
$1/|{\bf k}|$ potentials. However, in general it is the operators in
Eq.~(\ref{Ok1}) which are the fundamental quantities as discussed in
Sec.\,\ref{sect_usoft}.  The counterterms required to cancel the 
divergences in the two-body ultrasoft graphs in
Fig.~\ref{fig_usk}a,b,c plus the corresponding 
soft divergences determined by the pull-up mechanism~\cite{amis3} are
\begin{eqnarray} \label{dVk1}
 \delta {\cal V}_{k1}^{(1)} = -\frac{2C_A C_1}{3\pi}\Big[
 \frac{\alpha_s(\mu_U)}{\epsilon} - \frac{\alpha_s(\mu_S)}{\epsilon}\Big]
  \,,\quad
 \delta {\cal V}_{k2}^{(T)} = \frac{C_A(C_A+C_d)}{6\pi}\Big[
 \frac{\alpha_s(\mu_U)}{\epsilon} - \frac{\alpha_s(\mu_S)}{\epsilon}\Big]
  \,.
\end{eqnarray}
 
\begin{figure}[t!]
\centerline{ 
\includegraphics[width=1.4in]{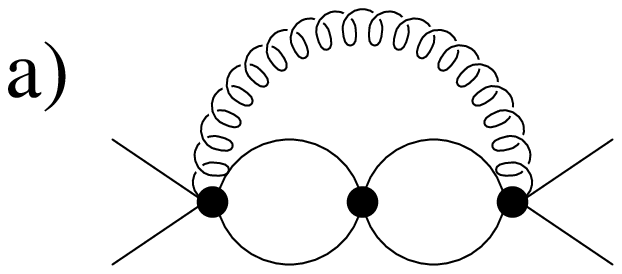} \hspace{0.5cm}
\includegraphics[width=1.4in]{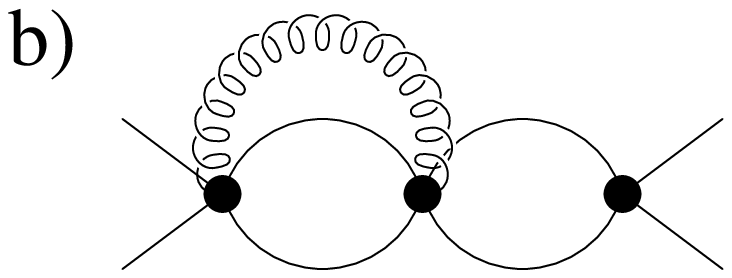} \hspace{0.5cm}
\includegraphics[width=1.3in]{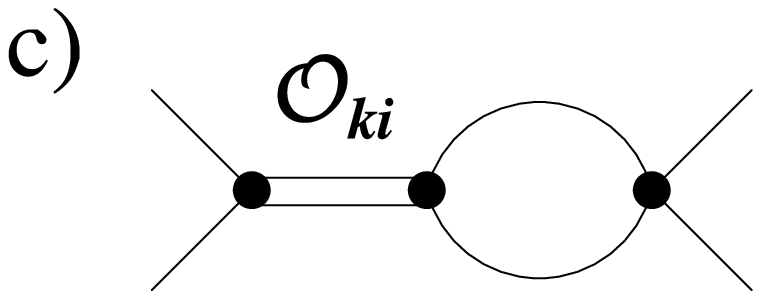} \hspace{0.3cm}
\includegraphics[width=1.3in]{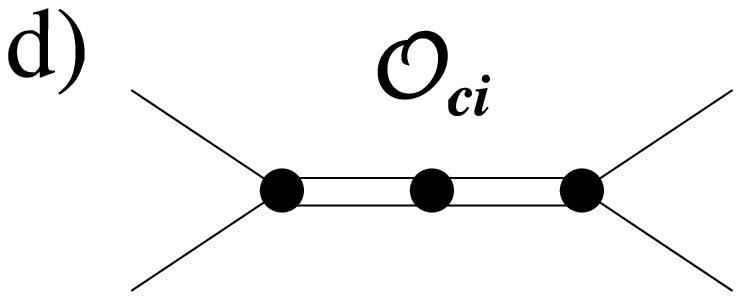} \hspace{0.2cm}  }
\vspace{-0cm} \caption{Graphs for the renormalization of the operator which 
leads to the $1/{\bf k}^2$ potential.
\label{fig_usc}}
\end{figure}

Next consider the order $\alpha^4 v^{-1}$ ultrasoft graphs. At this order there
are two diagrams (shown in Figs.~\ref{fig_usc}a,b) with ultrasoft gluons which
contain divergences that are not canceled by counterterms from ${\cal
V}_{2,r,k1,k2}$~\cite{hms1}. A part of the UV divergences in Fig.~\ref{fig_usc}b
is canceled by a $\delta {\cal V}_{k1,k2}$ counterterm graph involving the
function $f_2$ as shown in Fig.~\ref{fig_usc}c. Similar to the graphs in
Fig.~\ref{fig_usk} the remaining divergences are subtracted by operators with
sums over ${\bf q}, {\bf q'}$ as pictured in Fig.~\ref{fig_usk}d. The required
operators are
\begin{eqnarray} \label{Oc1}
  {\cal O}_{c1}^{(1)} &=& - [\mu_S^{2\epsilon}\,{\cal V}_c^{(T)}]^3
   \sum_{{\bf p,p',q,q'}}   (2h_0-h_1)\:
    \big[ \psi_{\bmp^\prime}^\dagger \psi_{\bmp}
   \chi_{-\bmp^\prime}^\dagger \chi_{-\bmp} \big] \,,
 \nn\\
   {\cal O}_{c2}^{(T)} &=& -[\mu_S^{2\epsilon}\,{\cal V}_c^{(T)}]^3 
   \sum_{{\bf p,p',q,q'}} h_0\: \big[ \psi_{\bmp^\prime}^\dagger T^A \psi_{\bmp}
   \chi_{-\bmp^\prime}^\dagger \bar T^A \chi_{-\bmp} \big] \,,
 \nn\\
   {\cal O}_{c3}^{(T)} &=& -[\mu_S^{2\epsilon}\,{\cal V}_c^{(T)}]^3
   \sum_{{\bf p,p',q,q'}}  h_1 \: \big[\psi_{\bmp^\prime}^\dagger T^A \psi_{\bmp}
   \chi_{-\bmp^\prime}^\dagger \bar T^A \chi_{-\bmp} \big]
 \,,
\end{eqnarray}
where ${\cal V}_{c1}^{(1)}$, ${\cal V}_{c2}^{(T)}$, ${\cal V}_{c3}^{(T)}$ are
the corresponding Wilson coefficients that appear in the resulting contribution
to the vNRQCD Lagrangian, $\Delta {\cal L}_p = {\cal V}_{c1}^{(1)} {\cal
O}_{c1}^{(1)} + {\cal V}_{c2}^{(T)} {\cal O}_{c2}^{(T)} + {\cal V}_{c3}^{(T)}
{\cal O}_{c3}^{(T)}$. The $\mu_S^\epsilon$ factors in Eq.~(\ref{Oc1}) are
determined as discussed in sections~\ref{sect_formalism},\ref{sect_usoft}.  The
functions $h_{0,1}$ have the form
\begin{eqnarray}
  h_0 &=& \frac{(\bmq^\prime-\bmp^\prime)\cdot (\bmq-\bmp)}
   {(\bmq-\bmp)^4(\bmq-\bmq^\prime)^2(\bmq^\prime-\bmp^\prime)^4} 
   \,, \quad \nn\\
  h_1 &=& \frac{(\bmq-\bmq^\prime)\cdot (\bmq-\bmp)}
   {(\bmq-\bmp)^4(\bmq-\bmq^\prime)^4(\bmq^\prime-\bmp^\prime)^2} 
   + (\bmp \leftrightarrow \bmp^\prime,\bmq\leftrightarrow \bmq^\prime )\,.
\end{eqnarray}
The counterterms needed to cancel the ultrasoft divergences in
Figs.~\ref{fig_usc}a,b,c plus the soft divergences determined by the
pull-up~\cite{hms1} are
\begin{eqnarray} \label{dVc1}
 \delta {\cal V}_{c1}^{(1)} &=& \frac{C_A C_1(C_A+C_d)}{12\pi}\Big[
 \frac{\alpha_s(\mu_U)}{\epsilon} - \frac{\alpha_s(\mu_S)}{\epsilon}\Big] 
  \,, \nn\\ 
 \delta {\cal V}_{c2}^{(T)} &=& \frac{4C_A C_1}{3\pi}\Big[
 \frac{\alpha_s(\mu_U)}{\epsilon} - \frac{\alpha_s(\mu_S)}{\epsilon}\Big] 
  \,, \nn\\  
 \delta {\cal V}_{c3}^{(T)} &=& \frac{2C_A}{3\pi}
  \Big[C_1+\frac{(C_A+C_d)^2}{32}\Big] \Big[
 \frac{\alpha_s(\mu_U)}{\epsilon} - \frac{\alpha_s(\mu_S)}{\epsilon}\Big] 
  \,.
\end{eqnarray}

From the counterterms in Eqs.~(\ref{dVk1}) and (\ref{dVc1}) the vRGE gives the
anomalous dimensions [$\gamma_{u2}(\nu) \equiv 2\alpha_s(m\nu^2) -
\alpha_s(m\nu)$]
\begin{eqnarray}
  \nu \frac{\partial}{\partial\nu} {\cal V}_{k1}^{(1)} &=& 
  -\frac{4C_A C_1}{3\pi} \gamma_{u2}(\nu)
  \,, \qquad\qquad\qquad 
 \nu \frac{\partial}{\partial\nu}  {\cal V}_{k2}^{(T)} = 
  \frac{C_A(C_A+C_d)}{3\pi}\gamma_{u2}(\nu)  \,,\nn \\[3pt]
  \nu \frac{\partial}{\partial\nu} {\cal V}_{c1}^{(1)} &=& 
  \frac{C_A C_1(C_A+C_d)}{6\pi} \gamma_{u2}(\nu)
  \,, \quad\qquad\! 
 \nu \frac{\partial}{\partial\nu}  {\cal V}_{c2}^{(T)} = 
  \frac{8 C_A C_1}{3\pi}\gamma_{u2}(\nu)  \,, \nn\\[3pt]
 \nu \frac{\partial}{\partial\nu}  {\cal V}_{c3}^{(T)} &=& 
  \frac{4C_A}{3\pi} \Big[C_1+\frac{(C_A+C_d)^2}{32}\Big] \gamma_{u2}(\nu)  \,.
\end{eqnarray}
Note that possible soft graphs which could contribute to these anomalous
dimensions are exactly canceled by ${\cal V}_c^{(T)}$ counterterms similar to
the QED example in Sec.\,\ref{sect_usoft}.  Using the boundary
conditions ${\cal V}_{ki}^{(1,T)}(1) = {\cal V}_{ci}^{(1,T)}(1) = 0$
we find the solutions 
\begin{eqnarray} \label{Vk1Vc1}
  {\cal V}_{k1}^{(1)}(\nu) &=& \frac{8C_A C_1}{3\beta_0}\: \ln(w)\,,
  \qquad\qquad\qquad\qquad
  {\cal V}_{k2}^{(T)}(\nu) = -\frac{2C_A(C_A+C_d)}{3\beta_0}\: \ln(w)\,, 
   \nn\\[3pt]
  {\cal V}_{c1}^{(1)}(\nu) &=& -\frac{C_A C_1(C_A+C_d)}{3\beta_0}\:\ln(w)\,,
    \qquad\qquad
  {\cal V}_{c2}^{(T)}(\nu) = -\frac{16C_A C_1}{3\beta_0}\: \ln(w)\,, 
   \nn\\[3pt]  
  {\cal V}_{c3}^{(T)}(\nu) &=& -\frac{8C_A}{3\beta_0}
     \bigg[ C_1 + \frac{(C_A+C_d)^2}{32} \bigg] \ln(w) \,,
\end{eqnarray}
where $w$ is given in Eq.~(\ref{w}).  These are our final results for the Wilson
coefficients of the new operators in Eqs.~(\ref{Ok1}) and (\ref{Oc1}). The
corresponding color singlet channel coefficients are given by ${\cal
V}^{(s)}={\cal V}^{(1)}-C_F{\cal V}^{(T)}$ (where coefficients for color
structures not shown in Eqs.~(\ref{Ok1}) and (\ref{Oc1}) such as ${\cal
V}_{k1}^{(T)}$ are zero at this order).

With the operators in Eqs.~(\ref{Ok1}) and (\ref{Oc1}) all divergences in the
usoft diagrams in Figs.\,~\ref{fig_usk}a,b,c and \ref{fig_usc}a,b,c are
canceled completely. Therefore, the corresponding contributions in the
anomalous dimensions for the potential coefficients ${\cal V}^{(1,T)}_{k,c}$ in
Eq.~(\ref{vNRQCDpotential}) obtained in Refs.~\cite{amis3,hms1} should be
removed.  The only remaining divergences that must be canceled by ${\cal
V}^{(1,T)}_{k,c}$ are from purely soft diagrams and are associated with the
known running of the strong coupling $\alpha_s$. Thus we find
\begin{eqnarray} \label{totad}
\nu \frac{\partial}{ \partial\nu} {\cal V}_k^{(T)}(\nu) &=&
  - \frac{\beta_0}{8\pi}  (7 C_A-C_d) \: {[\alpha_s(m\nu)]^3} \,, \qquad
\nu \frac{\partial }{ \partial\nu} {\cal V}_k^{(1)}(\nu) =
   - \frac{\beta_0}{2\pi}  \,C_1\: {[\alpha_s(m\nu)]^3} \nn \,,\\[5pt]
\nu \frac{\partial }{ \partial\nu} {\cal V}_c^{(T)}(\nu) &=& 
  -2  \bigg[ \beta_0 \alpha_s^2(m\nu) 
    +\beta_1 \frac{\alpha_s^3(m\nu)}{4\pi}
    +\beta_2 \frac{\alpha_s^4(m\nu)}{(4\pi)^2} \bigg]\,,\qquad  
\end{eqnarray}
where $\beta_{0,1,2}$ are the coefficients of the QCD beta function
(in the $\overline{\rm MS}$ scheme for $\beta_2$). With the matching 
conditions at $\nu=1$~\cite{amis2} these equations give the
solutions 
\begin{eqnarray} \label{VkVc}
  {\cal V}_k^{(T)}(\nu) = \frac{(7C_A\!-\!C_d)}{8}\: \alpha_s^2(m\nu)\,,\quad
  {\cal V}_k^{(1)}(\nu) = \frac{C_1}{2}\: \alpha_s^2(m\nu)\,,\quad
  {\cal V}_c^{(T)}(\nu) = 4\pi \alpha_s^{[3]}(m\nu) \,.
\end{eqnarray} 
Here $\alpha_s^{[3]}(m\nu)$ is the QCD coupling with 3-loop running.

It must be noted that in general the operators ${\cal O}_{kj}$ and ${\cal
O}_{cj}$, are {\em not} directly related to the potentials $1/|{\bf k}|$ and
$1/{\bf k^2}$ in Eq.~(\ref{vNRQCDpotential}). The reason is that the sums over
${\bf q,q'}$ must be regulated in the same way as sums over ${\bf p,p'}$. This
has important implications for the cancelation of subdivergences as discussed
in Sec.\,~\ref{sect_usoft} and therefore also affects renormalized matrix
elements as we will discuss further in the next two sections. However, if we
take {\em finite} matrix elements of the $1/|{\bf k}|$ and $1/{\bf k^2}$
potentials and operators ${\cal O}_{kj}$ and ${\cal O}_{cj}$, then we can use
$\sum_{\bf q} f_0 = \sum_{\bf q} f_1/3 =\sum_{\bf q} f_2 = 1/(16|{\bf k}|)$ and
$\sum_{\bf q,q'} h_0 =\sum_{\bf q,q'} h_1/2 = 1/(64\pi^2 {\bf k}^2)$. The result
for the finite matrix element is then equivalent to the matrix element of
effective $1/|{\bf k}|$ and $1/{\bf k^2}$ potentials with modified Wilson
coefficients:
\begin{eqnarray} \label{V1teff}
  {\cal V}_{k,{\rm eff}}^{(1)} &=& {\cal V}_k^{(1)}(\nu) 
    + {6\,\alpha_s^2(m\nu)}\: {\cal V}_{k1}^{(1)}(\nu)\,,\qquad
  {\cal V}_{k,{\rm eff}}^{(T)} = {\cal V}_k^{(T)}(\nu)
    + {4\,\alpha_s^2(m\nu)}\: {\cal V}_{k2}^{(T)}(\nu)\,,\nn\\[3mm]
  {\cal V}_{c,{\rm eff}}^{(1)} &=& {\cal V}_c^{(1)}(\nu) \,,\qquad
  {\cal V}_{c,{\rm eff}}^{(T)} = {\cal V}_c^{(T)}(\nu)
    + {\pi\,\alpha_s^3(m\nu)}\: \big[{\cal V}_{c2}^{(T)}(\nu)
 	+2 {\cal V}_{c3}^{(T)}(\nu) \big]\,.
\end{eqnarray}
Substituting in the solutions in Eqs.~(\ref{Vk1Vc1}) and (\ref{VkVc}) this
gives the following effective color singlet coefficients 
\begin{eqnarray} \label{Vseff}
  {\cal V}_{k,{\rm eff}}^{(s)} &=& 
    \frac{C_F}{2} (C_F\!-\!2 C_A)\,\alpha_s^2(m\nu)
    + \frac{8 C_F C_A (C_A+2 C_F)}{3\beta_0}\, \alpha_s^2(m\nu)\ln(w)\,,\nn\\
  {\cal V}_{c,{\rm eff}}^{(s)} &=& -4\pi C_F \alpha_s^{[3]}(m\nu)
    + \frac{2\pi\,C_F C_A^3}{3\beta_0}\, \alpha_s^3(m\nu)\:\ln(w) \,.
\end{eqnarray}
We emphasize that these effective coefficients can only be used for finite
matrix elements, and in this case they should be viewed simply as a shorthand
way of accounting for insertions of the ${\cal O}_{ki}$ and ${\cal O}_{ci}$
operators and the $1/|\bmk|$ and $1/\bmk^2$ potentials.  The results for these
effective coefficients agree with the results for the coefficients of the
$1/|{\bf k}|$ and $1/{\bf k}^2$ potentials in Refs.~\cite{PSstat,Pineda1} for
scales $\mu<mv$ (pNRQCD), but only {\em if} we impose the scale correlation
$\nu_{us}=m\nu^2$, and $1/r=m\nu$ as mentioned previously in
Sec.\,\ref{sectionQCDm2}. Again, our results do not agree with
Refs.~\cite{PSstat,Pineda1} for scales $mv <\mu< m$ (\mNRQCD\!\!).  Furthermore,
except for finite matrix elements, our results disagree with
Refs.~\cite{amis3,hms1,PSstat,Pineda1} since we have found that the anomalous
dimensions are associated to a different set of operators which includes ${\cal
O}_{k1}$, ${\cal O}_{k2}$, ${\cal O}_{c1}$, ${\cal O}_{c2}$, and ${\cal O}_{c3}$
rather than just the $1/|{\bf k}|$ and $1/{\bf k}^2$ potentials. In general
these operators contribute in a different way for matrix elements with
divergences.

\section{Results for the production current} \label{sectionc1}

The leading order production currents in the effective theory are of
order $v^3$ and produce a $q\bar q$ pair in a ${}^3S_1$ or ${}^1S_0$
state: 
\begin{eqnarray} \label{J1J0}
  {\bf J}_{1,\bf p} = 
    \psi_{\bmp}^\dagger\, \bsigma (i\sigma_2) \chi_{-\bmp}^*
   \,,\qquad
  J_{0,\bf p}=   \psi_{\bmp}^\dagger\,  (i\sigma_2) \chi_{-\bmp}^*\,.
\end{eqnarray}
where $c_1(\nu)$ and $c_0(\nu)$ are the corresponding Wilson coefficients. The
fields are evaluated at spacetime coordinate $x$.  The two-loop matching for
$c_1(1)$ and $c_0(1)$ were first considered in Refs.~\cite{firstc1,firstc0}, and
are scheme dependent.  With the potentials used here the matching onto $c_1(1)$
is known in the $\overline {\rm MS}$ scheme at two-loop
order~\cite{hmst,hmst1}. This matching condition is not affected by our new soft
operators ${\cal O}_{2i}^{(2)}$ or the potential operators ${\cal O}_{ki,ci}$
since their Wilson coefficients vanish identically at the hard scale.

The currents in Eq.~(\ref{J1J0}) receive a non-trivial NLL anomalous dimension
from graphs starting at order $\alpha_s^2\, v^0$ which were computed in
Ref.~\cite{LMR}. However, this anomalous dimension is affected by our ${\cal
O}_{ki}$ operators which were not included there. The additional contribution is
\begin{figure}[t!]
\centerline{ 
\includegraphics[width=2in]{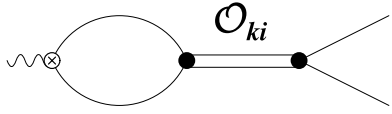}
 }
\caption{Contribution to the anomalous dimension for the production current
from ${\cal O}_{k1}^{(1)}$, ${\cal O}_{k2}^{(T)}$.
\label{c1k}}
\end{figure}
through the graph in Fig.~\ref{c1k} and we find the result
\begin{eqnarray} \label{c1ad}
\gamma_{c_1}(\nu) = \nu \frac{\partial}{\partial\nu} \ln[c_1(\nu)] &=&
 -\frac{{\cal V}_c^{(s)}(\nu)
  }{ 16\pi^2} \bigg[ \frac{ {\cal V}_c^{(s)}(\nu) }{4 }
  +{\cal V}_2^{(s)}(\nu)+{\cal V}_r^{(s)}(\nu)
   + {\bf S}^2\: {\cal V}_s^{(s)}(\nu)  \bigg] 
   \nn\\
  && + \frac{1}{2} {\cal V}_{k}^{(s)}(\nu) +\: \alpha_s^2(m\nu)\,\big[
   3 {\cal V}_{k1}^{(s)}(\nu) + 2 {\cal V}_{k2}^{(s)}(\nu) \big] \,,
\end{eqnarray}
where ${\bf S}^2=2$ for this spin-triplet coefficient. The anomalous dimension
for the spin singlet coefficient $c_0$ is identical to $\gamma_{c_1}(\nu)$, but
with ${\bf S}^2=0$.  The last term in Eq.~(\ref{c1ad}) which is proportional to
$\alpha_s^2(m\nu)$ is the contribution from Fig.~\ref{c1k}. However, it is easy
to see from Eq.~(\ref{V1teff}) that in this case the sum of the last two terms
is simply equal to $V_{k,{\rm eff}}^{(s)}(\nu)/2$ so that
\begin{eqnarray} \label{c1ad2}
 \gamma_{c_1}(\nu) &=&
 -\frac{{\cal V}_c^{(s)}(\nu)
  }{ 16\pi^2} \bigg[ \frac{ {\cal V}_c^{(s)}(\nu) }{4 }
  +{\cal V}_2^{(s)}(\nu)+{\cal V}_r^{(s)}(\nu)
   + {\bf S}^2\: {\cal V}_s^{(s)}(\nu)  \bigg] 
   + \frac{1}{2} {\cal V}_{k,{\rm eff}}^{(s)}(\nu) \,.
\end{eqnarray}
To see how this comes about note that the counterterms for the graph involving
the $1/|{\bf k}|$ potentials and the one in Fig.~\ref{c1k} are respectively
\begin{eqnarray} \label{deltac1}
  \delta c_1 &=& \frac{1}{4\epsilon} \: {\cal V}_k^{(s)}(\nu) \,,\qquad\\[3pt]
  \delta c_1 &=& \frac{1}{8\,\epsilon}\:  \Big[ 6 {\cal V}_{k1}^{(1)}(\nu)
  -4 C_F {\cal V}_{k2}^{(T)}(\nu) \Big] \alpha_s^2(m\nu) 
  = \frac{1}{8\,\epsilon}\: \big[ {\cal V}_{k,{\rm eff}}^{(s)}(\nu)
  -{\cal V}_k^{(s)}(\nu) \big]\,. \nn
\end{eqnarray}
However, the factor of two difference is made up for by the fact that there is
an additional relative factor of two in determining the anomalous dimensions
$\nu \partial/\partial\nu\, {\cal V}_k^{(s)} = -2\epsilon\, {\cal
V}_k^{(s)}+\ldots$, and $\nu \partial/\partial\nu [\alpha_s^2 {\cal
V}_{ki}^{(s)}] = -4\epsilon [\alpha_s^2 {\cal V}_{ki}^{(s)}]+\ldots$ in
dimensional regularization (to be compared with $\mu\partial/\partial\mu\,
\alpha_s = -2\epsilon\,\alpha_s+\ldots$). This difference stems from the 
factors $\mu_S^{2\epsilon}$ and $\mu_S^{4\epsilon}$ in Eqs.\,(\ref{Lp}) and
(\ref{Ok1}), respectively.  In the next section we will consider an example
where the ${\cal V}_{k1,k2}$ coefficients do not come in the linear combination
in ${\cal V}_{k,{\rm eff}}$.

Solving Eq.~(\ref{c1ad}) we find [$z=\alpha_s(m\nu)/\alpha_s(m)$,
$w=\alpha_s(m\nu^2)/\alpha_s(m\nu)$]
\begin{eqnarray} \label{c1sln}
 \ln\Big[ \frac{c_1(\nu)}{c_1(1)} \Big] &=&
  + a_2\,\pi\alpha_s(m)\,\left(1-z\right) 
  + a_3\, \pi \alpha_s(m) \ln(z) 
\nn\\ &&
 + a_4\, \pi\alpha_s(m) \Big[\,1- z^{1-13C_A/(6\beta_0)} \,\Big]
 + a_5 \, \pi\alpha_s(m) \Big[\,1- z^{1-2 C_A/\beta_0} \, \Big] 
\nn \\  && 
+ a_0\, \pi\alpha_s(m) \Big[\, (z-1)-w^{-1}\ln(w) \,\Big] \,,
\end{eqnarray}
where the coefficients $a_4$ and $a_5$ agree with Ref.~\cite{amis3}
\begin{eqnarray} \label{acoeffs2}
  a_4 &=& \frac{24 C_F^2 (11 C_A\!-\!3\beta_0)(5 C_A\!+\!8 C_F) }{ 13\, C_A
     (6\beta_0\!-\!13 C_A)^2},\quad
  a_5 = \frac{C_F^2 \big[ C_A( 15\! -\! 14\,{\bf S^2} ) 
    \!+\! \beta_0 (4{\bf S^2}\!-\!3) \big] }{ 6 (\beta_0\!-\!2 C_A)^2 }\,,
\end{eqnarray}
where ${\bf S}^2=2$. The modifications to the running potentials appearing in
Eq.~(\ref{c1ad}) cause the remaining terms to differ
\begin{eqnarray} \label{acoeffs}
  a_2 &=& 
   \frac{ C_F [ C_A\,C_F ( 9 C_A - 100 C_F) 
   - \beta_0\,( 26 C_A^2 + 19 C_A C_F - 32 C_F^2 ) ]}{ 26\,\beta_0^2\, C_A }
   \,, \nn\\
  a_3 &=&
    \frac{C_F^2 }{ \beta_0^2\, (6\beta_0-13C_A) (\beta_0-2C_A)}
    \, \Big\{ C_A^2 ( 9 C_A- 100 C_F )
    + \beta_0\,C_A \Big[ 74 C_F + C_A ( 13 {\bf S^2} - 42 ) \Big]\nn \\
   &&\qquad - 6 \beta_0^2  \Big[ 2 C_F + C_A ( {\bf S^2} - 3 ) \Big]\,\Big\} 
   \,,\nn\\[5pt]
  a_0 & = & -\frac{8\,C_F\,(C_A+C_F)\,(C_A+2\,C_F)}{3\beta_0^2} \,.
\end{eqnarray}
For the solution for $c_0(\nu)$ one should substitute ${\bf S}^2=0$ in the $a_i$
coefficients.

It is important to note that the anomalous dimension $\gamma_{c_1}(\nu)$ in
Eqs.\,(\ref{c1ad},\ref{c1ad2}) arises from divergences in potential loop
diagrams which must be computed with a dynamic fermion propagator $i/[E-{\bf
p}^2/(2m)]$~\cite{LMR}.  In Ref.~\cite{amis3} it was shown that this anomalous
dimension correctly reproduces the $\alpha_s^3\ln^2\alpha_s$ terms in the
wavefunction at the origin first computed in
Ref.~\cite{Penin}.\footnote{
This result is not changed by our results here since the coefficient
of the first log in the resummed ${\cal V}^{(s)}_j$ coefficients is
the same as in Ref.~\cite{amis3} and 
$\int d\nu/\nu \ln\nu = (\ln^2\nu)/2$.
}  For this to be the case it was necessary to include both
soft and ultrasoft contributions to the running potentials on the right hand
side of Eq.~(\ref{c1ad}). As the scale for this anomalous dimension travels from
$m$ to $mv$ it was necessary to simultaneously have the soft loop contributions
vary from $m$ to $mv$ at the same rate, and have the ultrasoft loops vary from
$m$ to $mv^2$.  Furthermore, in the Appendix of Ref.\,\cite{hmst1} it was found
that the relation $\mu_U=\mu_S^2/m$ is required for a consistent subtraction of
subdivergences coming from ultrasoft and potential divergences in three-loop
vertex diagrams that contribute to the NNLL anomalous dimension of the
production currents. This shows that ultrasoft gluons are needed starting at the
scale $m$. It also shows that it is necessary to simultaneously have divergent
loops with ultrasoft, soft and potential momentum in the theory.

We believe that for dynamic quarks these facts are difficult to reconcile from
the point of view of the \mpNRQCD formalism. The existence of \mpNRQCD seems to
depend crucially on there being a non-trivial stage of matching that occurs at a
scale $\mu\simeq mv\simeq 1/r$ where soft gluons are integrated out, while the
above results seem to indicate that such an intermediate matching scale does not
exist in general. This does not mean that results obtained with the \mpNRQCD
formalism are necessarily wrong for dynamic quarks, but seems to imply that they
may require some reinterpretation.

In Ref.\,\cite{Pineda2} a procedure was developed which reproduces the anomalous
dimension in Eq.~(\ref{c1ad2}) in the framework of the \mpNRQCD formalism. In
our opinion, the procedure suggested in Ref.\,\cite{Pineda2} contradicts some
features of \mpNRQCD upon which other results seem to rely. In particular, in
Ref.~\cite{Pineda2} it was proposed to a) demand a correlation between the
energy and momentum scales but only for the $c_1$ computation, b) transport the
$1/r$ matching scale back up to $m$ by hand so that ultrasoft gluons exist for
all scales $\mu<m$, and c) allow couplings associated with soft loops to become
unfrozen and run down again from $m$.  With this construction the potentials
used in Ref.~\cite{Pineda2} on the RHS of Eq.~(\ref{c1ad2}) agree with ours, so
the solution in Eq.~(\ref{c1ad}) agrees with the one in Ref.~\cite{Pineda2}
setting $\nu_p=m\nu$. For this to be the case it was crucial that it was the
combination ${\cal V}_{k,{\rm eff}}^{(s)}$ that appeared in Eq.~(\ref{c1ad2}),
and that the corresponding contributions in Ref.~\cite{Pineda2} were treated in
the same way as they are predicted to be treated by vNRQCD.  We agree with
Ref.~\cite{Pineda2} that static quarks can be used to simplify certain
calculations, in particular those with soft loops. However, our conclusion is
then that mNRQCD does not exist by itself as a physical theory that can be used
to make predictions with dynamic quarks.

Our results for the running of $c_1(\nu)$ and the ${\cal V}_j^{(s)}(\nu)$ have
implications for the case where $\Lambda_{\rm QCD}$ becomes comparable to
$mv^2$. Take $mv^2\sim \Lambda_{\rm QCD}$ and consider decreasing $\nu$ from
$\nu=1$. As the cutoff on momentum transfers and quark momenta, $m\nu$, gets
close to $mv$, the $\mu_U$ scale in the ultrasoft $\alpha_s(\mu_U)$ couplings is
approaching the scale $\Lambda_{\rm QCD}$.  Since ultrasoft gluons renormalize
the potentials these effects are tied together as soon as ultrasoft gluons first
start to renormalize operators. Thus, as $m\nu$ gets close to $mv$, a scale
affecting the coefficients ${\cal V}_j(\nu)$ of the potentials approachs
$\Lambda_{\rm QCD}$.\footnote{The evolution towards the scale $\Lambda_{\rm
QCD}$ due to ultrasoft gluons can still be computed perturbatively. This is very
much like the fact that the evolution for the region $\Lambda_{\rm QCD}<\mu<m_c$
in $B\to D$ decays can be computed perturbatively~\cite{Falk}. One must just be
careful not to evolve too close to the scale $\Lambda_{\rm QCD}$.} Considering
one ultrasoft gluon the affected potentials include the spin-independent $1/m^2$
potentials, and effective $1/|{\bf k}|$ and $1/{\bf k}^2$ potentials (or more
properly the ${\cal O}_{ci}$ and ${\cal O}_{ki}$ operators).  Furthermore, due
to the correlation between $E$ and ${\bf p}^2/(2m)$ for the heavy quarks,
non-perturbative effects at the energy scale could lead to non-perturbative
effects for the momenta. One conclusion is that for $mv^2\sim \Lambda_{\rm QCD}$
we can become sensitive to non-perturbative scales through the potentials even
for cutoffs near the momentum transfer $mv\gg \Lambda_{\rm QCD}$. This seems
quite problematic for perturbatively matching onto the potentials at a scale
$\mu=mv$ for dynamic quarks. However, this does not affect the matching for
static quarks as done with the \mpNRQCD formalism in Ref.~\cite{Brambilla}.  For
$b\bar b$ states with dynamic quarks and $mv^2\sim\Lambda_{\rm QCD}$, this might
also imply that non-perturbative effects have a larger influence than one
usually infers.

\section{Top Production at Threshold and Quarkonium Energies}
\label{sectiontop}

The results for the order $v^{-1}$, $v^0$ and $v^1$ QCD potentials and the new
operators ${\cal O}_{ki}$ and ${\cal O}_{ci}$ presented in the previous sections
affect the results for the NLL and NNLL top threshold $e^+e^-$ cross
section that were given in Refs.\,\cite{hmst,hmst1} and the NNLL
quarkonium energies given in 
Ref.~\cite{hms1}.  Except for the handling of the ${\cal O}_{ki}$ and $1/|{\bf
k}|$ potentials discussed below, the changes all involve simply substituting in
the coefficients ${\cal V}_{2}^{(s)}$ and ${\cal V}_r^{(s)}$ given in
Eq.~(\ref{pv2}) and replacing ${\cal V}_{c}^{(s)}$ by the ${\cal V}_{c,{\rm
eff}}^{(s)}$ given in Eq.~(\ref{Vseff}).

The modified results for the coefficients of the order $v^1$ spin-independent
potentials, ${\cal V}_{2}^{(s)}$ and ${\cal V}_r^{(s)}$, and the
production current, $c_1$, affect the results only
trivially through the modified running of the coefficients. They do not lead to
any change in the analytic form of the NLL and NNLL corrections to the current
correlator ${\cal A}_1$.  For the corrections caused by the $v^{-1}$ (Coulomb)
potential and the operators ${\cal O}_{ci}$ it is sufficient to consider the
effective Coulomb potential coefficient in Eq.\,(\ref{Vseff}) because the
corresponding matrix elements do not lead to UV divergences which affect the
cross section.  Thus, in the results presented in Ref.\,\cite{hmst,hmst1} one
simply has to be replace ${\cal V}_c^{(s)}\to$ ${\cal V}_{c,{\rm eff}}^{(s)}$
and there is no other change.  In Table.~\ref{Vnums} we compare results for the
coefficients at $\nu=0.15$ and $\nu=0.3$. The changes are smaller for the larger
value of $\nu$ as expected. For $\nu=0.15$ we see that the changes in ${\cal
V}_2^{(s)}$, ${\cal V}_c^{(s)}$ and $c_1(\nu)$ are still very small
($1.1\%$, $0.08\%$ and $0.2\%$, respectively), while ${\cal V}_r^{(s)}$
changes by a more moderate amount 
($8\%$). For completeness $V_{k,{\rm eff}}^{(s)}$ is also shown, even though it
is not just this combination of coefficients that appears in the cross section.
Since $V_{k,{\rm eff}}^{(s)}(\nu)$ has a zero near $\nu\simeq 0.15$ the relative
change for this value of $\nu$ is quite large. A more relevant measure is the
suppression of ${\cal V}_{k,{\rm eff}}^{(s)}$ at $\nu\simeq 0.15$ compared to at
$\nu=1$, which is observed in both results. The effect of these
changes on the cross section are discussed later in this section.
\begin{table}[t!] 
\begin{center}
  \begin{tabular}{lccccc}
   &	\hspace{0.35cm}$\,{\cal V}_r^{(s)}(\nu)$\hspace{0.35cm} 
   & 	\hspace{0.35cm} $\,{\cal V}_2^{(s)}(\nu)$ \hspace{0.35cm} 
   &  	\hspace{0.35cm} $\pi^2 {\cal V}_{k,{\rm eff}}^{(s)}(\nu)$ \hspace{0.35cm}
   &	\hspace{0.35cm} $\,{\cal V}_{c,{\rm eff}}^{(s)}(\nu)$ \hspace{0.35cm} 
   &	\hspace{0.35cm} $\,c_1(\nu)/c_1(1)$ \hspace{0.35cm} 
   \\ \hline
  \mbox{matching $(\nu=1)$} 
	&  $-1.81$	&  $0$		&  $-0.357$	&  $-1.800$ 
        &  $1.000$ \\[2mm]
  \mbox{old results $(\nu=.3)$}
	&  $-1.72$  &  $0.361$	&  $-0.256$	&  $-2.153$ 
        &  $1.034$ \\
  \mbox{new results $(\nu=.3)$}   
	&  $-1.68$  &  $0.359$	&  $-0.238$	&  $-2.153$ 
        &  $1.034$ \\[2mm]
  \mbox{old results $(\nu=.15)$}
	&  $-1.51$  &  $0.616$	&  $-0.043$	&  $-2.423$ 
        &  $1.046$ \\
  \mbox{new results $(\nu=.15)$}   
	&  $-1.39$  &  $0.609$	&  $\phantom{-}0.016$	&  $-2.425$
        &  $1.044$ 
 \end{tabular}
\end{center}
\caption{Comparison of the numerical change to the Wilson coefficients for
$t\bar t$ ($m_t=175\,{\rm GeV}$). The old results show the coefficients used in
Ref.~\cite{hmst1}, while the new results use the coefficients from
Sects.~\ref{sectionQCDm2} and \ref{sectionQCDm0m1}. For the first
three columns and the fourth column
we use 1-loop running for $\alpha_s$, while the last column uses 3-loop
running. The coefficient ${\cal V}_s^{(s)}(\nu)$ is unchanged from the result in
Ref.~\cite{amis} and is not shown.\label{Vnums}}
\end{table}

The new results for the $v^0$ potentials and the operators ${\cal O}_{ki}$ lead
to analytic changes in the NNLL corrections to the current correlator ${\cal
A}_1$ because the corresponding matrix elements are UV divergent and the
n-dependent contributions that arise from summing the potential indices in the
operators ${\cal O}_{ki}$ in dimensional regularization lead to modifications of
the UV-finite terms.  With dimensional regularization the full n-dependent
($n=d-1$) expression for the potential that appears in the Schr\"odinger
equation from the operators ${\cal O}_{ki}$ (and ${\cal O}_{ci}$) is given in
App.\,\ref{appendix2}.  Altogether, there are now three different types of
corrections to the current correlator $\delta G^k$, $\delta G^{k1}$, and $\delta
G^{k2}$, that are needed to account for the corrections originating from the
singlet $1/|\bmk|$ potential and the operators ${\cal O}_{k1}$ and ${\cal
O}_{k2}$, respectively. Using $\overline{\rm MS}$ we find
\begin{eqnarray} \label{Gks}
 \delta G^{k1}(a,v,m,\nu) 
 & = & 
 -\frac{3 m^2}{4 \pi\, a}\left\{\,
 i\,v - a\left[\,\ln\left(\frac{-i\,v}{\nu}\right)
 -\frac{17}{12}+\ln 2+\gamma_E+\Psi\left(1-\frac{i\,a}{2\,v}\right)\,\right]
 \,\right\}^2
\label{deltaGkn1}
\nn \\ 
\lefteqn{
 +\,\frac{3m^2}{4 \pi\,a}\,\left[\, -v^2 + 
 \frac{a^2}{16}\,\left(\frac{1}{\epsilon^2}
 -\frac{11}{3\epsilon} -\frac{89}{9}\right)
 \,\right]
\,,
}
 \nonumber \\[3mm]
 \delta G^{k2}(a,v,m,\nu) 
 & = & 
 -\frac{m^2}{2 \pi\, a}\left\{\,
 i\,v - a\left[\,\ln\left(\frac{-i\,v}{\nu}\right)
 -\frac{21}{16}+\ln 2+\gamma_E+\Psi\left(1-\frac{i\,a}{2\,v}\right)\,\right]
 \,\right\}^2
\label{deltaGkn2}
\nn \\ 
\lefteqn{
 +\,\frac{m^2}{2 \pi\,a}\,\left[\, -v^2 + 
 \frac{a^2}{16}\,\left(\frac{1}{\epsilon^2}
 -\frac{13}{4\,\epsilon} - \frac{175}{16}\right)
 \,\right]
\,,
}
\end{eqnarray}
In deriving these equations we have included the counterterm generated by
renormalizing the ${\bf J}_1$ current at NLL order in Eq.~(\ref{deltac1}). These
counterterm graphs are sufficient to cancel all subdivergences. The remaining
overall divergences shown in Eq.~(\ref{Gks}) are canceled by counterterms for
the current correlator. For comparison the term $\delta G^k$ from the $1/|{\bf
k}|$ potential which was given in Ref.\,\cite{hmst1} is:
\begin{eqnarray}
 \delta G^k(a,v,m,\nu) 
 & = & 
 -\frac{m^2}{8 \pi\, a}\left\{\,
 i\,v - a\left[\,\ln\left(\frac{-i\,v}{\nu}\right)
 -2+2\ln 2+\gamma_E+\Psi\left(1-\frac{i\,a}{2\,v}\right)\,\right]
 \,\right\}^2
\label{deltaGk} 
\nn \\
\lefteqn{ \hspace{-1.4cm}
 +\,\frac{m^2}{8 \pi\,a}\,\left[\, -v^2 + 
 \frac{a^2}{4}\,\left(\frac{1}{3\,\epsilon^2}
 -\frac{2}{\epsilon}\left(1-\frac{2}{3}\ln 2\right)
 + \frac{4}{3}-8\ln 2+\frac{8}{3}\ln^2 2+\frac{\pi^2}{9}\right)
 \,\right],
}
\end{eqnarray}
where this expression is also renormalized in the way described above. Since the
$\delta G^{ki}$ are not proportional to $\delta G^k$, it is in general
{\em not} the linear combination of coefficients defined in ${\cal
  V}_{k,{\rm eff}}^{(s)}$ in 
Eq.~(\ref{Vseff}) that appears in the cross section. The NNLL vector correlator
\begin{eqnarray}
 {\cal A}_1(v,m,\nu) & = & i\,
 \sum\limits_{\mbox{\scriptsize\boldmath $p$},\mbox{\scriptsize\boldmath $p'$}}
 \int\! d^4x\: e^{i \hat{q} \cdot x}\:
 \Big\langle\,0\,\Big|\, T\, {\bf J}_{1,\bf p}(x)\,{\bf J}_{1,\bf p'}^\dagger(0)
 \Big|\,0\,\Big\rangle 
\end{eqnarray}
can be expressed in terms of the renormalized Greens functions as
\begin{eqnarray} \label{A1}
{\cal A}_1
 & = &
 6 \,N_c\,\bigg[\,
 G^c(a',v,m,\nu)
 + \left({\cal{V}}_2^{(s)}(\nu)+2{\cal{V}}_s^{(s)}(\nu)\right)\, 
 \delta G^\delta(a,v,m,\nu)
+ \,{\cal{V}}_r^{(s)}(\nu)\,\delta G^r(a,v,m,\nu) 
\nonumber
\\[2mm] & &
+ \,{\cal{V}}_k^{(s)}(\nu)\, \delta G^k(a,v,m,\nu) 
  +  \alpha_s^2(m\nu)\,{\cal{V}}_{k1}^{(1)}(\nu)\, \delta G^{k1}(a,v,m,\nu)
\nonumber
\\[2mm] & &
  -  C_F\,\alpha_s^2(m\nu)\,{\cal{V}}_{k2}^{(T)}(\nu)\, \delta G^{k2}(a,v,m,\nu)
+ \,\delta G^{\rm kin}(a,v,m,\nu) 
\,\bigg]
\,,
\label{NNLLcrosssection}
\\[2mm]
a & = & -\frac{1}{4\,\pi}\,{\cal V}_{c}^{(s)}(\nu)\,,
\qquad
a' =  -\frac{1}{4\,\pi}\,{\cal V}_{c,{\rm eff}}^{(s)}(\nu)
\,,
\end{eqnarray}
where $\hat q=(\sqrt{s}-2m,{\bf 0})$, $m$ is the top quark pole mass,
$v=(\sqrt{s}-2m+i\Gamma_t)/m)^{1/2}$, and the expressions for $G^c(a,v,m,\nu)$,
and $\delta G^{\delta,r,{\rm kin}}(a,v,m,\nu)$ are given in
Ref.\,\cite{hmst1}. The combination ${\rm Im}[c_1^2(\nu){\cal A}_1]$ then
appears in $R^v$, which is the normalized vector current induced cross
section. From analyzing Eq.~(\ref{A1}) we see that while the coefficients of
some functions are proportional to ${\cal V}_{k,{\rm eff}}^{(s)}$, the
coefficients of other terms are not. For instance we have terms such as ${\cal
V}_{k,{\rm eff}}^{(s)}\: \Psi^2(z)$, however the combination $(34{\cal
V}_{k1}^{(s)} +21{\cal V}_{k2}^{(s)})\Psi(z)$ also appears. Therefore, in
general the effective $1/|{\bf k}|$ potential is not sufficient and the
individual expressions for the ${\cal V}_{k,k1,k2}(\nu)$ contributions are
required.

Our results for the ${\cal V}^{(s)}_j$ coefficients also numerically affect the
NNLL relation between the 1S mass and the pole mass, which is needed to switch
to the 1S mass scheme~\cite{Andre1S}. This relation is obtained from the
${}^3S_1$ ground state solution of the Schr\"odinger equation and is given in
Eq.\,(46) of Ref.\,\cite{hmst1}. The modifications caused by the order $v^1$
potentials are again trivial, and only the coefficients ${\cal V}_r^{(s)}$ and
${\cal V}_2^{(s)}$are changed. For the NNLL energy levels the matrix elements of
the ${\cal O}_{ci}$ and ${\cal O}_{ki}$ operators do not cause additional UV
divergences, so the effect of these operators can be implemented simply by
replacing ${\cal V}_{c}^{(s)}$ and ${\cal V}_{k}^{(s)}$ in Ref.\,\cite{hmst1} by
${\cal V}_{c,{\rm eff}}^{(s)}$ and ${\cal V}_{k,{\rm eff}}^{(s)}$ respectively.
Thus, the additional correction to the current correlator ${\cal A}_1$ caused by
switching to the 1S mass scheme has the same functional form as before
\begin{eqnarray}
\delta {\cal A}_1(v,M^{\rm 1S},\nu)
& = &
-\,6 \,N_c\, \frac{\Delta_m^{\mbox{\tiny NNLL}}}{v}\,
\frac{d}{dv}\,G^0(a,v,M^{\rm 1S},\nu)
\label{deltaA1S}
\\[2mm]
\Delta_m^{\mbox{\tiny NNLL}} & = &
 - \frac{a^2}{8}\,
  {\cal V}_{k,{\rm eff}}^{(s)}
 -\,\frac{a^3}{8\pi}\,\bigg(\frac{{\cal V}_2^{(s)}}{2} + 
 {\cal V}_s^{(s)} + \frac{3 {\cal V}_r^{(s)}}{8}\bigg) 
 + \frac{5}{128}\,a^4 
\,,\hspace{5mm}
\nonumber
\end{eqnarray}
where $G^0$ is the LL zero-distance Greens function of the Schr\"odinger
equation given in Eq.~(34) of Ref.~\cite{hmst1}.  The formula for the heavy
quarkonium spectrum for arbitrary quantum numbers is also affected by simply
replacing coefficients in Eq.~(40) of Ref.\,\cite{hms1} just as in the
expression for the ${}^3S_1$ ground state energy described above.

Next we turn to how our results numerically affect the vector-current-induced
top threshold cross section $R^v$ at NLL and NNLL order. We will see
that the change is quite small and well within the error estimate of
Ref.~\cite{hmst1}.  Note that the axial-vector cross section $R^a$
does not receive any modifications as it depends only on the LL value
of ${\cal V}_c^{(s)}$ at this order.

\begin{figure}[t!]
\begin{minipage}{2.9in}
\begin{tabular}{cll|rrrr}
 & $\sqrt{s}$ (GeV) & & $347$ & $350$ & $353$ & \\ 
 \hline
 & $Q_t^2 R^v_{\rm LL}$ & 
  $\nu=.1\ \ $  &\: $0.386$ &\: $1.556$ &\: $1.276$ \\
 & \hspace{0.9cm}\raisebox{0.2cm}{} &  
  $\nu=.125$\,\,& $0.354$  & $1.411$ & $1.215$  \\ 
 & \hspace{0.9cm}\raisebox{0.2cm}{} &  
  $\nu=.275$    & $0.276$  & $1.054$ & $1.043$  \\ \hline
 & $Q_t^2 R^v_{\rm NLL}$ & 
  $\nu=.1\ \ $  & $0.235$ & $0.900$ & $0.788$ \\
 & \hspace{0.9cm}\raisebox{0.2cm}{} &  
  $\nu=.125$    & $0.240$  & $0.930$ & $0.815$  \\ 
 & \hspace{0.9cm}\raisebox{0.2cm}{} &  
  $\nu=.275$    & $0.243$  & $0.939$ & $0.840$  \\ \hline
 & $Q_t^2 R^v_{\rm NNLL}$ & 
  $\nu=.1\ \ $  & $0.241$ & $0.902$ & $0.858$ \\
 & \hspace{0.9cm}\raisebox{0.2cm}{} &  
  $\nu=.125$    & $0.242$  & $0.926$ & $0.844$  \\ 
 & \hspace{0.9cm}\raisebox{0.2cm}{} &  
  $\nu=.275$    & $0.245$  & $0.960$ & $0.845$ 
\end{tabular}
\end{minipage}\hspace{0.2cm}
\begin{minipage}{3.4in}
  \hbox{\includegraphics[width=3.4in]{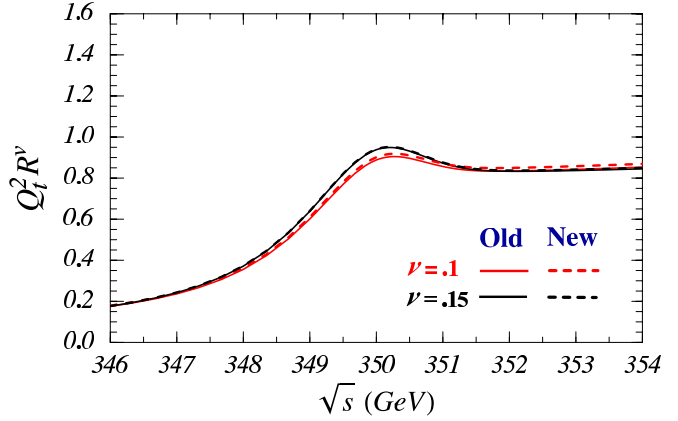}}
\end{minipage}
\caption{Updated numerical values of $Q_t R^v$ in the 1S mass
scheme. Also shown is a figure comparing the old NNLL results from
Ref.~\cite{hmst1} (solid curves) and our new results (dashed curves) for two
values of $\nu$ (red $\nu=0.1$, and black $\nu=0.15$). Near the peak it is easy
to see that the scale uncertainty is several times larger than the difference
between the old and new results.
\label{tabm1Svalues}}
\end{figure}

%
%
%
\begin{figure}[t] 
\begin{center}
\leavevmode
\epsfxsize=3.8cm
\leavevmode
\epsffile[210 585 408 710]{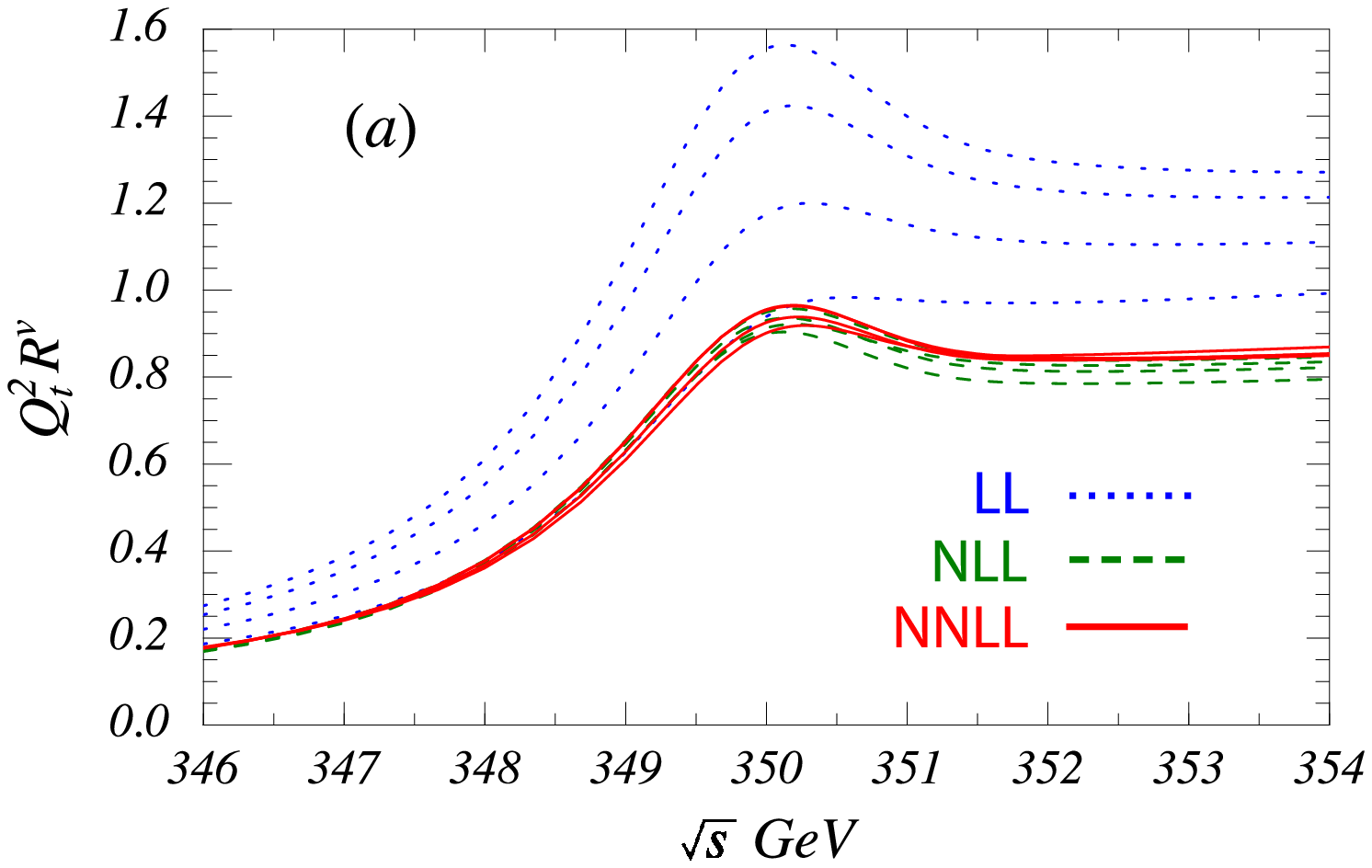}
\hspace{4.3cm}
\leavevmode
\epsfxsize=3.8cm
\leavevmode
\epsffile[220 580 420 710]{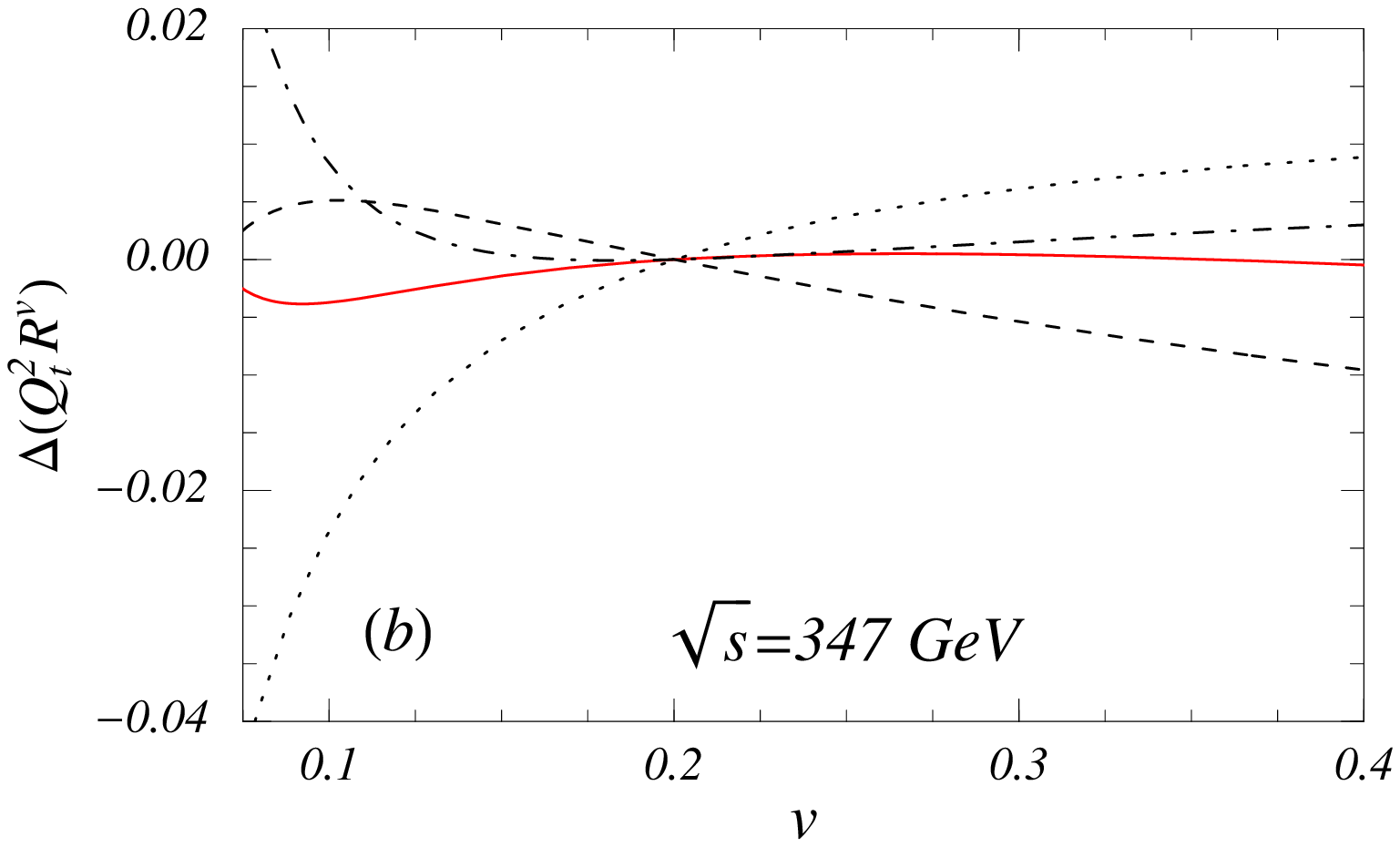}
\\[3.2cm]
\leavevmode
\epsfxsize=3.8cm
\leavevmode
\epsffile[220 580 420 710]{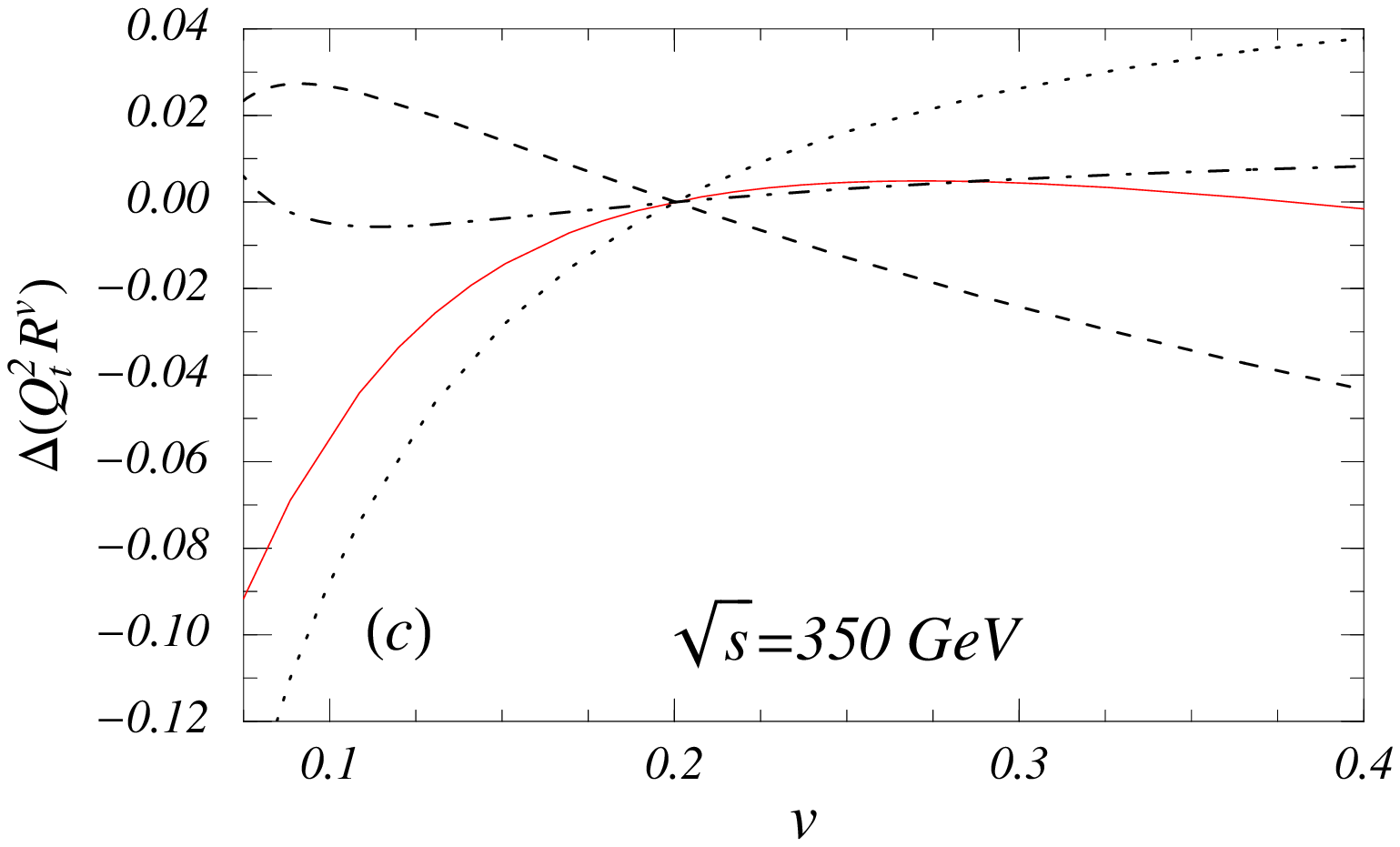}
\hspace{4.3cm}
\leavevmode
\epsfxsize=3.8cm
\leavevmode
\epsffile[220 580 420 710]{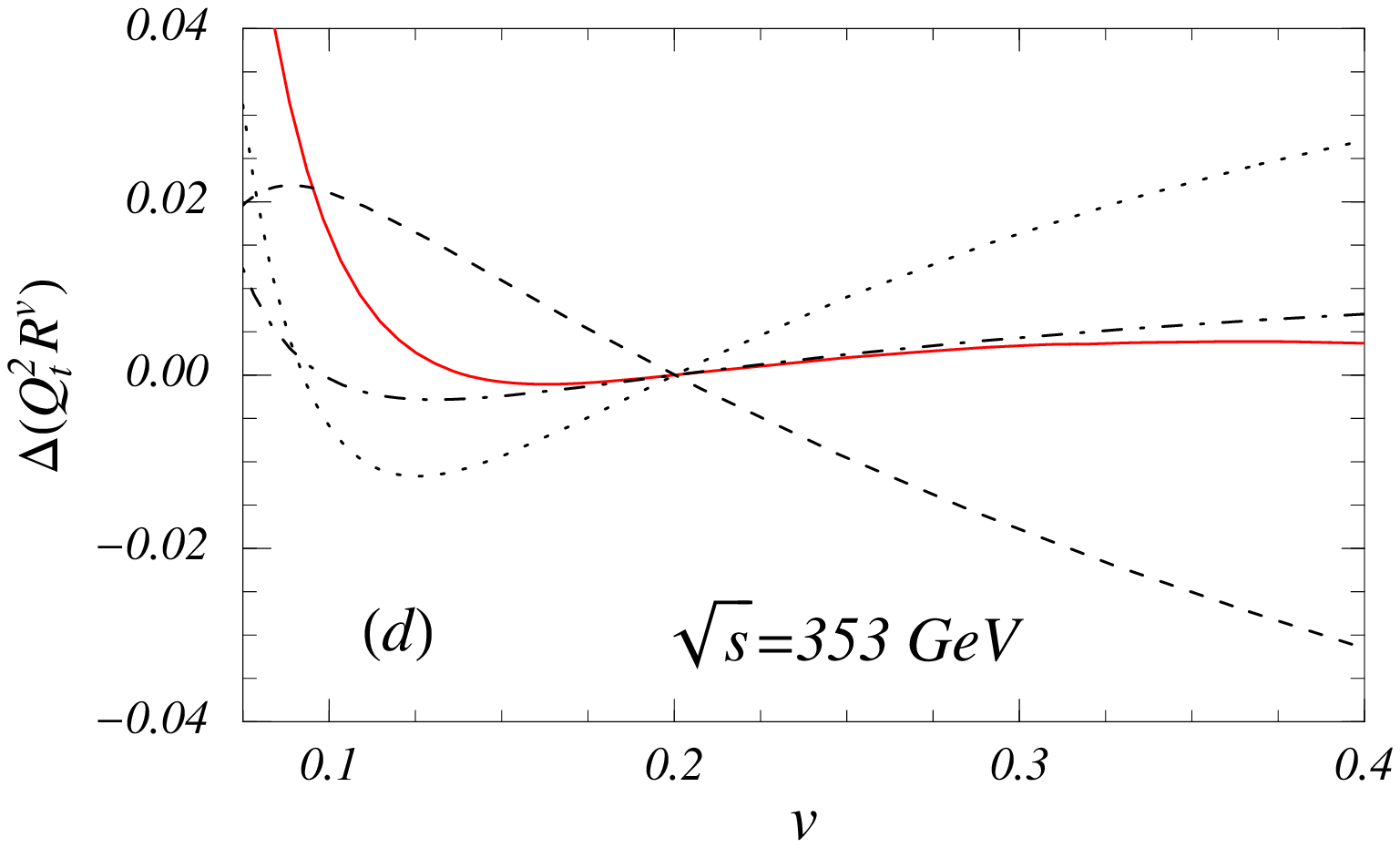}
\vskip  3.cm
 \caption{ Panel a) shows the NNLL results for $Q_t^2 R^v$ with $M^{\rm
 1S}=175$~GeV at LL (dotted lines), NLL (dashed lines) and NNLL (solid lines)
 order. For each order four curves are plotted for $\nu=0.1$, $0.125$, $0.2$ and
 $0.4$. Panels b)-d) show the relative scale dependence of various NNLL
 contributions to $Q_t^2 R^v$ for different c.m.\,energies. The contributions
 are divided into those from $G^c$ (dashed lines), the sum off $\delta
 G^{k,k1,k2}$ (dotted line), and the sum of $\delta G^{\delta,r,{\rm kin,1S}}$
 (dot-dashed lines) while the solid (red) lines denote the sum of these terms,
 $Q_t R^v$. Note that the plots have different scales for the y-axes.
\label{figtopplots} 
}
 \end{center}
\end{figure}

In Fig.\,\ref{tabm1Svalues} we have displayed our updated values of $Q_t^2 R^v$
in the 1S mass scheme as a function of the c.m.\.energy $\sqrt{s}$ for $M^{\rm
1S}=175$~GeV, $\alpha_s(m_Z)=0.118$ and $\Gamma_t=1.43$~GeV, and taking
four-loop running for $\alpha_s$ with $n_f=5$ active massless quark flavors. The
numerical methods used to evaluate $G_c$ were described in
Ref.~\cite{Jezabek}. We also show a figure comparing our NNLL results to those
in Ref.~\cite{hmst1} for $\nu=0.1$ and $\nu=0.15$.  The relative deviation is
quite small and essentially independent of the c.m.\,energy at NLL and NNLL
order. Note that this is obvious for the NLL cross section, because at this
order only the coefficient $c_1$ is affected by the new results. For
$\sqrt{s}=350\,{\rm GeV}$ the relative deviation amounts to $(2.2\%,
1.0\%,1.0\%, 0.6\%, 0.3\%, 0.1\%)$ at NLL order and to $(1.5\%, 0.6\%,0.2\%,
0.01\%, 0.1\%, 0.1\%)$ at NNLL order for
$\nu=(0.1,0.125,0.15,0.2,0.275,0.4)$. This is well within the relative
uncertainty of $\pm 3\%$ for the normalization of $Q_t^2 R^v$ at NNLL order
which we estimated in Ref.\,\cite{hmst1}.

In Fig.\,\ref{figtopplots}a we have displayed $Q_t^2 R^v$ up to NNLL order.  The
curves show the LL (dotted glue lines), NLL (dashed green lines) and NNLL (solid
red lines) cross section for $\nu=0.1, 0.125, 0.2$ and $0.4$, where at LL order
lower curves correspond to larger values of $\nu$.  The conclusions drawn from
these results are the same as in Ref.~\cite{hmst1}. Thus, compared to earlier
NNLO (fixed order) results~\cite{Hoang3} the variation of the normalization of
the NNLL cross section with $\nu$ is considerably reduced. Equally important,
the sum of all NNLL corrections is about an order of magnitude smaller than the
size of the NNLO corrections~\cite{Hoang3}, and agrees well with the predictions
at NLL order. This indicates that the expansion is converging.

In Figs.\,\ref{figtopplots}b, c and d we have displayed separately, the
contributions to $Q_t^2 R^v$ coming from $G^c$ (dashed lines), from the sum of
$\delta G^{k,k1,k2}$ (dotted lines) and the sum of $\delta G^{\delta,r,{\rm
kin,1S}}$ (dot-dashed lines) as a function of $\nu$ for $\sqrt{s}=347, 350$ and
$353$~GeV. In general we find for $\nu\gsim 0.15$ that the $\nu$-variation of
the contributions from $G^c$ and from the sum of $\delta G^{k,k1,k2}$ cancel to
a large extent, whereas the contributions from $\delta G^{\delta,r,{\rm
kin,1S}}$ are almost $\nu$-independent. On the other hand, for values of $\nu <
0.15$ and energies around and above the peak position the $\nu$-variation is
dominated by the contributions from $\delta G^{k,k1,k2}$, which rapidly increase
in size for decreasing $\nu$, whereas the contributions from $G^c$ and $\delta
G^{\delta,r,{\rm kin,1S}}$ are small. The behavior of these results are in
agreement with our previous results in Refs.\,\cite{hmst,hmst1} and shows that
the modified results for the potentials and the effects of the new operators
${\cal O}_{ci}$ and ${\cal O}_{ki}$ lead only to small numerical changes while
the essential properties remain unchanged.  To be more specific, making a
conservative estimate by using the value $0.1$ as the lower bound for the
velocity scaling parameter~\cite{hmst1} to determine the remaining theoretical
uncertainties of the NNLL renormalization group improved cross section we find
\begin{equation}
  \frac{\delta\sigma_{t\bar t}}{\sigma_{t\bar t}} \, = \, \pm 3\,\% \,.
\end{equation}
This agrees with our earlier estimate in Ref.\,\cite{hmst1} and is an
order of magnitude smaller than the uncertainties associated to 
fixed order NNLO QCD computations~\cite{Hoang3}. In particular, the
conclusions that we have drawn in Ref.\,\cite{hmst1} concerning the
theoretical uncertainties in extractions of $\alpha_s$, the top Yukawa
coupling and the total top width from a threshold scan at a future
linear collider remain unchanged and are comparable to the expected
experimental uncertainties~\cite{TTbarsim}.

\section{Conclusion} \label{section_conclusion}

In this paper we have reconsidered the renormalization group improvement of
Wilson coefficients in NRQCD in light of two new observations about the
structure of ultrasoft renormalization of operators.  

We first showed that the ultrasoft renormalization of operators with soft gluons
can induce through mixing operators ${\cal O}_{2i}^{(2)}$ whose Wilson
coefficients vanish at the matching scale. Taking four quark matrix elements of
these operators then causes a renormalization of the spin-independent $1/m^2$
potentials.  Using the notation for the $1/m^2$ potential in
Eq.~(\ref{vNRQCDpotential}) this analysis affects the running of the
coefficients ${\cal V}_2(\nu)$ and ${\cal V}_r(\nu)$ in QCD.  Our results are
different from Ref.~\cite{amis} where these additional operators were not
included. We also compared our results to those in the \mpNRQCD approach in
Ref.~\cite{Pineda1}. Our results do not agree with the running of 4-quark
operators in \mNRQCD because we find that the renormalization from ultrasoft
gluons is present for all scales $\mu<m$. We did find agreement with the pNRQCD
results if we imposed a correlation between the \mpNRQCD matching and
renormalization scales, $\nu_{us}=m\nu^2$ and $1/r=m\nu$. This agreement is
encouraging, however our correlation requirement may have implications for the
NRQCD analysis of the case $mv^2\sim\Lambda_{\rm QCD}$, as discussed in
Sec.\,~\ref{sectionc1} and mentioned below.

Second we performed an analysis of graphs containing mixed ultrasoft and
potential loops and gave a new procedure for subtracting divergences in these
graphs. In particular we find that ultrasoft gluons renormalize the operators
${\cal O}_{ki}$ and ${\cal O}_{ci}$ displayed in Eqs.~(\ref{Ok1}) and
(\ref{Oc1}), rather than the $1/|{\bf k}|$ and $1/{\bf k}^2$ potentials. Because
of this our results differ from those in Refs.~\cite{amis3,hms1,PSstat,Pineda1}.
In certain situations with finite matrix elements it is possible to make use of
effective $1/|{\bf k}|$ and $1/{\bf k}^2$ potentials with modified Wilson
coefficients. However in general this is not the case. An example of the former
are predictions for the NNLL perturbative quarkonium energy levels, while an
example of the latter are current correlators which give the $e^+e^-\to t\bar t$
cross sections as discussed in Sec.\,~\ref{sectiontop}.

We also considered the implications of the modified running for the evolution of
the production current (Sec.\,~\ref{sectionc1}) and for predictions for the
$t\bar t$ cross section at NNLL order. For the cross section the change from the
results in Ref.~\cite{hmst,hmst1} is very small, being $\lesssim 1\%$ for the
physically motivated values $\nu>0.15$. If we include a larger more conservative
range of scale variations then the change is still small becoming $1.5\%$ at
$\nu=0.1$.  Since the uncertainty assigned to predictions in Ref.~\cite{hmst1}
was $\pm 3\%$, all analyses and conclusions for the cross section are unchanged.
In particular, there is still a large improvement in the convergence of the
perturbation theory over not summing the logarithms, and we can assign a
conservative overall uncertainty of $\pm 3\%$ to the NNLL cross section
predictions which is much smaller than the uncertainty found at
NNLO~\cite{Hoang3}.

Our results for the running have implications for cases where $\Lambda_{\rm
QCD}$ becomes comparable to $mv^2$. Essentially, since ultrasoft gluons
renormalize spin-independent potential operators these potentials become
sensitive to the scale $\Lambda_{\rm QCD}$ at an earlier stage than might
otherwise be thought. The list of affected potentials includes the
spin-independent $1/m^2$ potentials and the ${\cal O}_{ci}$ and ${\cal O}_{ki}$
operators (which give effects often attributed to the $1/|{\bf k}|$ and $1/{\bf
k}^2$ potentials). Due to the correlation for $mv^2\sim \Lambda_{\rm QCD}$ we
are sensitive to non-perturbative scales through the potentials even though the
momentum transfer $mv\gg \Lambda_{\rm QCD}$. This seems problematic for
perturbatively matching onto the potentials at a scale $\mu=mv$ for dynamic
quarks, because at this scale the coefficients of the potentials are already
blowing up.

\begin{acknowledgments}
We would like to thank A.~Manohar for many useful discussions and gratefully
acknowledge communication with A.~Penin on the nature of the $n_f \alpha_s^5
\ln^2 \alpha$ corrections to the $1/{\bf k}^2$ potential. We also thank
A.~Manohar and I.~Rothstein for comments on the manuscript. We thank the
Benasque Center of Science, where part of this work was carried out, and
I.S. would like to thank the Max-Planck-Institut in Munich for their hospitality
while this work was completed. This work was supported in part by the Department
of Energy under the grant DE-FG03-00-ER-41132.
\end{acknowledgments}

\newpage
\appendix

\section{Summary of structures for six field operators}
\label{appendix1}

The structures appearing in the QCD operators ${\cal O}_{2\varphi}^{(0)}$,
${\cal O}_{2A}^{(0)}$, and ${\cal O}_{2c}^{(0)}$ include
\begin{eqnarray} 
\Gamma_\varphi^{(0)}
& = & 
\bigg[\, 
T^AT^B\,
\frac{(2q^0\gamma^0+{\bf k}\cdot{\mbox{\boldmath$\gamma$}})}
 {{\bf k}^2+2 {\bf k}\cdot {\bf q}} +
T^BT^A\,
 \frac{(2q^0\gamma^0-{\bf k}\cdot{\mbox{\boldmath $\gamma$}})}
 {{\bf k}^2-2 {\bf k}\cdot {\bf q}}
\,\bigg]\,\Big(Z_0^{(0)}\Big)^2
\,,
\nonumber\\[2mm]
\Gamma^{(0)}_{\varphi,\psi} & = & T^A
\,, \qquad\qquad 
\Gamma^{(0)}_{\varphi,\chi} \, = \, \bar T^B
\nonumber\\[4mm]
\Gamma_{A,\mu\nu}^{(0),CD}
& = & 
-\frac{1}{2}\,f^{AEC}\,f^{BED}\,
\Big[\,
U^{(0)}_{\mu\sigma}(q,\bmp,\bmp^\prime)\,
(U^{(0)})_\nu^{\,\,\,\,\sigma}(-q,-\bmp,-\bmp^\prime)\,
\Big]\,
\frac{1}{{\bf k}^2+2 {\bf k}\cdot {\bf q}}
\nonumber\\&&
- \, \frac{1}{2}\,f^{AED}\,f^{BEC}\,
\Big[\,
U^{(0)}_{\mu\sigma}(-q,\bmp,\bmp^\prime)\,
(U^{(0)})_\nu^{\,\,\,\,\sigma}(q,-\bmp,-\bmp^\prime)\,
\Big]\,
\frac{1}{{\bf k}^2-2 {\bf k}\cdot {\bf q}}
\,,
\nonumber\\[2mm]
\Gamma^{(0)}_{A,\psi} & = & T^A
\,, \qquad\qquad 
\Gamma^{(0)}_{A,\chi} \, = \, \bar T^B
\nonumber\\[4mm]
\Gamma_c^{(0),CD}
& = & 
\,
\bigg[\, 
\frac{f^{AEC}\,f^{BED}}{{\bf k}^2+2 {\bf k}\cdot {\bf q}} +
\frac{f^{AED}\,f^{BEC}}{{\bf k}^2-2 {\bf k}\cdot {\bf q}}
\,\bigg]\,\Big(Y_0^{(0)}\Big)^2
\,,
\nonumber\\[2mm]
\Gamma^{(0)}_{c,\psi} & = & T^A
\,, \qquad\qquad 
\Gamma^{(0)}_{c,\chi} \, = \, \bar T^B
\label{F}
\end{eqnarray}
where the $U_{\mu\nu}^{(0)}$, $Z_0^{(0)}$, and $Y_0^{(0)}$ coefficients can be
found in Ref.~\cite{amis}.  We have also made use of color contractions of these
structures including
\begin{eqnarray} \label{Gam1} 
\Gamma_\varphi^{(0),(1)} &=& \Gamma_\varphi^{(0),(T)}
= 
\frac{1}{2}\bigg[\,
\frac{(2q^0\gamma^0+{\bf k}\cdot{\mbox{\boldmath$\gamma$}})}
 {{\bf k}^2+2 {\bf k}\cdot {\bf q}} +
 \frac{(2q^0\gamma^0-{\bf k}\cdot{\mbox{\boldmath $\gamma$}})}
 {{\bf k}^2-2 {\bf k}\cdot {\bf q}}
\,\bigg]\,\Big(Z_0^{(0)}\Big)^2
\,,
\nonumber\\[2mm]
\Gamma_{\phi,\psi}^{(0),(T)} &=& T^A\,,\quad 
\Gamma_{\phi,\chi}^{(0),(T)} = \bar T^A\,,\quad
\Gamma_{\phi,\psi}^{(0),(1)} = 1\,,\quad
\Gamma_{\phi,\chi}^{(0),(1)} = 1\,,
\nonumber\\[2mm]
\Gamma_{A,\mu\nu}^{(0),(T)} &=& \Gamma_{A,\mu\nu}^{(0),(1)} 
=
-\frac{C_A}{2}\,\bigg\{\,\Big[\,
U^{(0)}_{\mu\sigma}(q,\bmp,\bmp^\prime)\,
(U^{(0)})_\nu^{\,\,\,\,\sigma}(-q,-\bmp,-\bmp^\prime)\,
\Big]\,
\frac{1}{{\bf k}^2+2 {\bf k}\cdot {\bf q}}
\nonumber\\&& \qquad\quad
+ \,
\Big[\,
U^{(0)}_{\mu\sigma}(-q,\bmp,\bmp^\prime)\,
(U^{(0)})_\nu^{\,\,\,\,\sigma}(q,-\bmp,-\bmp^\prime)\,
\Big]\,
\frac{1}{{\bf k}^2-2 {\bf k}\cdot {\bf q}}
\,\bigg\}
\,,
\nonumber\\[2mm]
\Gamma_{A,\psi}^{(0),(T)} &=& T^A\,,\quad 
\Gamma_{A,\chi}^{(0),(T)} = \bar T^A\,,\quad
\Gamma_{A,\psi}^{(0),(1)} = 1\,,\quad
\Gamma_{A,\chi}^{(0),(1)} = 1\,,
\nonumber\\[2mm]
\Gamma_{c,\psi}^{(0),(T)} &=& \Gamma_{c,\psi}^{(0),(1)}
= 
C_A\, \bigg[\, 
\frac{1}{{\bf k}^2+2 {\bf k}\cdot {\bf q}} +
\frac{1}{{\bf k}^2-2 {\bf k}\cdot {\bf q}}
\,\bigg]\,\Big(Y_0^{(0)}\Big)^2 \,,
\nonumber\\[2mm]
\Gamma_{c,\psi}^{(0),(T)} &=& T^A\,,\quad 
\Gamma_{c,\chi}^{(0),(T)} = \bar T^A\,,\quad
\Gamma_{c,\psi}^{(0),(1)} = 1\,,\quad
\Gamma_{c,\chi}^{(0),(1)} = 1\,.
\end{eqnarray}

\section{Form of potentials in the Schr\"odinger equation}
\label{appendix2}

When the $1/|{\bf k}|$ potential and the operators ${\cal O}_{k1}$ and ${\cal
O}_{k2}$ are included in the Schr\"odinger equation in dimensional
regularization the form of the potential is
\begin{eqnarray}
 V_k^{\rm Schr.}(\bmp,\bmq)
& = &
 \frac{\pi^2\,\mu_S^{2\epsilon}}{m |\bmk|}\, {\cal{V}}_k^{(s)}(\nu) 
+ 
 \frac{\pi^2\,\mu_S^{4\epsilon}}{m (\bmk^2)^{2-\frac{n}{2}}}\,
 \alpha_s^2(m\nu)\,{\cal V}_{k1}^{(1)}(\nu)\,
 \Big[ 8(n^2-5n-12)\,f(1,1)\Big]
\nonumber\\ & &
-
 \frac{\pi^2\,\mu_S^{4\epsilon}}{m (\bmk^2)^{2-\frac{n}{2}}}\,
 C_F\,\alpha_s^2(m\nu)\,{\cal V}_{k2}^{(T)}(\nu)\,
 \Big[ 4(n^2-3n+8)\,f(1,1)\Big]
\,,
\label{vkschroedinger}
\end{eqnarray}
while for the $1/{\bf k}^2$ potential and the operators ${\cal
O}_{c1}$, ${\cal O}_{c2}$, and ${\cal O}_{c3}$ the potential is
\begin{eqnarray}
 V_c^{\rm Schr.}(\bmp,\bmq)
& = &
 \frac{\mu_S^{2\epsilon}}{\bmk^2}\,{\cal V}_c^{(s)}(\nu)
+ 
 \frac{64\pi^3\mu_S^{6\epsilon}}{(\bmk^2)^{4-n}}\,
 C_F\,\alpha_s(mv)^3\,
\Big(\frac{1}{2}{\cal V}_{c2}^{(T)}+{\cal V}_{c3}^{(T)}\Big)
\nonumber\\[2mm] & &
\hspace{2cm}
\,\times\Big[ (4-n)\,f(1,2)\,f(3-\frac{n}{2},1)
\Big]
\,,
\label{vcschroedinger}
\end{eqnarray}
where
\begin{eqnarray}
f(a,b) 
& = &
\frac{
\Gamma(a+b-\frac{n}{2})\Gamma(\frac{n}{2}-a)\Gamma(\frac{n}{2}-b)
}{\Gamma(a)\Gamma(b)\Gamma(n-a-b)(4\pi)^{\frac{n}{2}}} \,.
\end{eqnarray}
In Eq.~(\ref{vcschroedinger}) the operator ${\cal O}_{c1}$ gives a vanishing
contribution.  Note that the functions $1/({\bf k}^2)^{2-n/2}$ and $1/({\bf
k}^2)^{4-n}$ are {\em not} the d-dimensional Fourier transform of a $1/r^2$ or
$1/r$ potential.  Also note that in another regularization scheme such as with a
cutoff, the potentials in Eqs.~(\ref{vkschroedinger}) and (\ref{vcschroedinger})
would take a different functional form, however the momentum dependence of the
${\cal V}_k$, ${\cal V}_{k1,k2}$, ${\cal V}_c$, and ${\cal V}_{c2,c3}$ terms
would still differ from each other.


\end{document}